\documentclass[prc,aps,floatfix,showpacs,twocolumn,nofootinbib]{revtex4-2}
\usepackage{dcolumn}
\usepackage{slashed}
\usepackage{bm}
\usepackage{color}
\usepackage{graphicx}
\usepackage{amssymb,amsmath}
\usepackage[utf8]{inputenc} 							% lettere accentate da tastiera
\usepackage{bbm}
\usepackage{simplewick}
\usepackage{float}
\usepackage{simplewick}

\def\tri{{{}^3{\rm H}}}
\def\het{{{}^3{\rm He}}}
\def\heq{{{}^4{\rm He}}}

\def\bmr{{\bm r}}

\def\bmp{{\bm p}}

\def\bmq{{\bm q}}

\def\n{\phantom{0}}

\def\jac{x}
\newcommand{\jacb}{{\bm x}}
\def\hypfi{\varphi}

\newcommand{\bmsi}{{\bm \sigma}}

\begin{document}
%%%%%%%%%%%%%%%%%%%%%%%%%%%%%%%%%%%%%%%%%%%%%%%%%%%%%%%%%%%%%%%%%%%%%%%%%%%%%%%%%%%%
\title{A new analysis of the ``hep'' S-factor and the ``hen'' cross section} 

\author{M. Viviani$^1$, A. Gnech$^{2,3}$, L.E. Marcucci$^{4,1}$, 
A. Kievsky$^1$, and L. Girlanda$^{5,6}$}

\affiliation{
$^1$\mbox{INFN-Pisa, I-56127, Pisa, Italy} \\
$^2$\mbox{Department of Physics, Old Dominion University, Norfolk, VA 23529, USA}\\
$^3$\mbox{Theory Center, Jefferson Lab, Newport News, VA 23606, USA}\\
$^4$\mbox{Department of Physics ``E. Fermi'', University of Pisa, I-56127 Pisa, Italy} \\  
$^5$\mbox{Department of Mathematics and Physics, University of Salento, I-73100 Lecce, Italy }\\
$^6$\mbox{INFN-Lecce, I-73100 Lecce, Italy }}
\date{\today}

\begin{abstract}
  We present a new accurate analysis of the $\het(p,e^+\nu_e)\heq$ (``hep'') reaction
  at astrophysical energies. The S-factor is computed using a state-of-the-art method
  to calculate the four-nucleon scattering and bound-state wave functions (the hyperspherical
  harmonic expansion), and by using nuclear interactions and accompanying
  electroweak nuclear currents obtained within the chiral effective field theory framework. 
  Our analysis includes a detailed examination of the theoretical uncertainties coming
  from two different sources: the truncation of the interaction and current chiral expansions,
  and the model dependence.  Our recommended final theoretical value for the hep S-factor at zero energy
  is  $S(0)=(8.7\pm 0.9)\times 10^{-20}$ keV b. We provide also the energy spectrum of the outgoing hep positrons which may be
  measured in future experiments.
  We include also an analysis of the ``sister'' reaction $\het(n,\gamma)\heq$ (``hen'') at low energies,
  showing that the  calculation well reproduce the total cross section from
  thermal energies to few MeV, validating our results on the hep reaction.   
\end{abstract}


\maketitle

\section{Introduction}

In the solar $pp$-I chain of reactions~\cite{SF3}, the process $\het(p,e^+\nu_e)\heq$ (the so-called ``hep'' reaction)
is one of the possible mechanisms which can burn $\het$ nuclei (produced via the ${}^2$H$(p,\gamma)\het$ process) and
form  $\heq$ nuclei. This reaction produces the most energetic solar neutrinos, with an endpoint energy of $18.8$ MeV.
Its rate, however, is suppressed by various orders of magnitude with respect to the ``standard'' 
final reaction of the $pp$-I chain, namely the  $\het(\het,2p)\heq$ reaction~\cite{SF3}. The reasons for
this suppression are (i) the Coulomb repulsion between $p$ and $\het$,
being the typical energy of the reaction in the astrophysical environment of the order of a few keV, and (ii) the fact
that the reaction is induced by the weak interaction. Furthermore, 
the leading one-body Gamow-Teller current operator cannot connect the main $S$-state components of the $p+\het$ and $\heq$
initial- and final-state wave functions. As a consequence,
the one-body axial current contribution is suppressed, and the reaction proceeds through the small components of the $\het$ and $\heq$ wave
functions. The suppression is even enhanced by the cancellations between the one- and two-body weak current contributions.
Therefore,  the hep reaction
cross section is too small to be measured in laboratory and only theoretical predictions are available, similarly to the $pp$ fusion.

The most recent studies of this reaction are those reported in Refs.~\cite{Marcucci:2000bh,Marcucci:2001dmc,Park:2002yp}.
In all these studies, the $p+\het$  and $\heq$ initial and final nuclear wave
functions were obtained with the correlated hyperspherical harmonics (CHH) variational method~\cite{Viviani:1994pm,Viviani:1998gr},
using the Argonne $v_{18}$ (AV18) two-nucleon potential~\cite{Wiringa:1994wb} augmented by
the Urbana IX (UIX) three-nucleon interaction~\cite{Pudliner:1995wk}.

Regarding the weak vector and axial transition operators, in Refs.~\cite{Marcucci:2000bh,Marcucci:2001dmc}
they were obtained within a phenomenological approach, similar to those used for the $pp$ capture of Ref.~\cite{Schiavilla:1998je}.
In particular, the two-body contributions were obtained from the exchange of families of mesons, constructed
 to be consistent with the AV18/UIX interaction. Other contributions
were present, as those arising from the excitation of intermediate $\Delta$-isobar degrees
of freedom. An unknown coupling constant, multiplying the nucleon-to-$\Delta$ axial transition operator,
was fixed to reproduce the Gamow-Teller matrix element (GTME) of
tritium $\beta$-decay, as already done for the $pp$ fusion in Ref.~\cite{Schiavilla:1998je}.

In the subsequent study of Ref.~\cite{Park:2002yp}, the calculations were performed
using the same CHH $p+\het$  and $\heq$ initial and final nuclear wave functions,
computed starting from
the AV18/UIX interaction as in Ref.~\cite{Marcucci:2000bh,Marcucci:2001dmc}, but the
nuclear weak current operators were obtained within the framework of chiral effective field theory ($\chi$EFT).
In this approach, there is an unknown coupling constant, $d_R$, related to a contact axial current
(in $\chi$EFT such coupling constants are called low-energy constants -- LECs). 
which is fixed to reproduce the experimental value of the GTME.
We refer to this study as a hybrid calculation (wave functions calculated from
phenomenological interactions, while currents derived within $\chi$EFT). The results of this latter study were found in agreement
with those presented in Refs.~\cite{Marcucci:2000bh,Marcucci:2001dmc}.

In this paper, we have reconsidered the study of the ``hep'' reaction but 
within a fully consistent $\chi$EFT approach. More precisely, our aim is threefold:
(i) we will use more accurate $\heq$ and $p+\het$ wave functions calculated with the (uncorrelated) hyperspherical
hamonics (HH) method; this method allows for
a better convergence of the quantity of interest and also the possibility to use non-local momentum-space interactions.
(ii) We will consider both nuclear interactions and weak currents derived from $\chi$EFT for the first time.  (iii) we will study at the same time the
related process $n(\het,\gamma)\heq$ (the ``hen'' process), for which good experimental data are available. Notice that our work is able to provide an answer to all the recommendations mentioned in Ref.~\cite{SF3} for a new analysis of the hep process.

In our investigation, we follow an approach similar to that employed to study other nuclear reactions that uses consistent chiral interaction and currents such as $pp$ fusion~\cite{Marcucci:2013tda}, muon captures on light
nuclei~\cite{Acharya:2018qzk,Bonilla:2022otm,Ceccarelli:2022cpz,Gnech:2023mvb,Marcucci:2011jm,Golak:2016zcw}, and other $3+1$ reactions similar to the one treated here~\cite{Girlanda:2010vm,Viviani:2021stx}.
It is important to warn the reader that the interactions and currents are as consistent as possible. So far no fully consistent $\chi$EFT interactions and currents are available up to the chiral order of the interactions we are using in this work. 
In our analysis we include also a detailed examination of the theoretical uncertainties coming
from the truncation of the interaction and current chiral expansion, and the model dependence similar to the one proposed in Ref.~\cite{Gnech:2023mvb}.
We also provide the energy spectrum of the outgoing positrons which may be
measured in future experiments, as, for example, SNO+~\cite{SNO:2021xpa}.

This manuscript is organized as follow. 
In Sec.~\ref{sec:theo}, a brief description of the theoretical formalism is given. Then, in Sec.~\ref{sec:res}
the results of the calculations are reported and discussed.
Finally, in the last section, the conclusions of this study are reported. Further details of the calculations are reported in the Appendix.

\section{Theoretical analysis}
\label{sec:theo}

This section is organized as follow. In the first subsection we present the interactions and currents
used in this work. In the second subsection the expressions for the $p+\het$ and $n+\het$ wave functions
are given. The wave functions are calculated by means of the HH method~\cite{Kievsky:2008es,10.3389/fphy.2020.00069,Viviani:2020gkm}
(details regarding the definition of the HH functions and how to compute bound and scattering state wave functions
are given in Appendix~A).
In Secs.~\ref{sec:sf} and~\ref{sec:xs}, we report the formulas to obtain the hep S-factor, the spectra of emitted positrons,
and the hen cross section.
Finally, in the last subsection,  we briefly recall the method proposed in Ref.~\cite{Epelbaum:2014efa} and used in this work to estimate
the theoretical error due to the truncation of the chiral expansion for a given observable.

\subsection{The interactions and currents}
\label{sec:models}

The interactions we use in the present calculation are of two types.
The first ones are 
developed within the framework of $\chi$FFT
by Entem, Machleidt and Nosyk (EMN) in Ref.~\cite{Entem:2017gor}.
These interactions are implemented in momentum space and are strongly
non-local. The degrees of freedom are pions and nucleons only.
For this interaction family all the orders up to
the next-to-next-to-next-to-next-to-leading-order (N4LO)
are available for three different cutoff values
$\Lambda\,$=$\,450$, $500$ and $550$ MeV. The LECs of these
interactions are fixed fitting the pion-nucleon ($\pi N$) and the nucleon-nucleon (NN) database up to $300$ MeV.
In the following we will use also the shortcuts LO for leading order, NLO for next-to-leading order,
N2LO for next-to-next-to-leading order, and N3LO for next-to-next-to-next-to-leading order.

The second family of interactions, the so-called Norfolk potentials (NV)~\cite{Piarulli:2014bda,Piarulli:2016vel}, are local interactions 
still derived in the framework of $\chi$EFT at N3LO,
and include $\Delta$-isobars together with pions and
nucleons as degrees of freedom. However, the final form has been obtained by
disregarding some non-local terms, plus all N3LO pion-exchange diagrams, so
they cannot be considered complete at this order. The interactions are regularized in
configuration-space with two regulators, one ($R_S$) for the short-range components associated with NN
contact terms, and the other ($R_L$) for the long-range terms, coming from pion exchanges.
We consider four different interactions of this family, for which two different sets of regulators have been used,
and the LECs have been fitted considering the NN database within two different energy ranges. In Table~\ref{tab:potlist} we  summarize the names and characteristics of the interactions used in this work.

We also include in the Hamiltonian a chiral 3N interaction, derived at N2LO
in Refs.~\cite{Epelbaum:2002vt}, containing the LECs $c_1$, $c_3$, and $c_4$, 
entering in particular in the  two-pion exchange (2PE) terms
(they also appear in the NN potential and in the weak current and are clearly
 fixed from NN and $\pi N$ scattering data), 
 and two additional parameters usually denoted $c_D$ and $c_E$,
 which have to be fixed using 3N data.
The LEC $c_D$ is linearly dependent on the LEC $d_R$ entering
the axial current at N3LO. Therefore, in our approach, we have fixed the LEC $c_D$ and $c_E$ by fitting simultaneously the binding energy of $^3$H and the GTME of the $^3$H $\beta$-decay (which depends on $d_R$). The fit has been performed using the axial current up to  N3LO for each interaction except LO and NLO EMN, for which the N3LO axial current term proportional to $d_R$ is not defined.
The adopted values of $c_D$ and $c_E$ obtained in the fitting procedure in Refs.~\cite{Baroni:2018fdn} and~\cite{Gnech:2023mvb} for the NV and EMN interaction, respectively, are reported in Table~\ref{tab:potlist}.

As we have stated, the N2LO 3N interaction depends also on
the LECs $c_{1,3,4}$. 
Regarding the N3LO and N4LO 3N 
forces~\cite{Bernard:2007sp,Bernard:2011zr,Krebs:2012yv,Krebs:2013kha,Girlanda:2011fh}, 
due to their complexity, we have neglected them.
However, in Ref.~\cite{Krebs:2012yv}, it has been
shown that the 2PE 3N force has essentially the same mathematical 
structure at N2LO, N3LO, and N4LO. Thus, one way to take into account the 2PE terms
at N3LO and N4LO is to use the N2LO expressions but with "effective" LECs $c_{1,3,4}$.
These LECs for the EMN potentials were provided in Ref.~\cite{Entem:2017gor}. 
In such a way, the EMN NN+3N interactions N3LO500/3N, N4LO450/3N, N4LO500/3N,
and N4LO550/3N (even using a N2LO 3N force)
can be (approximately) considered as ``full'' N3LO and N4LO interactions, respectively.
We refer to Ref.~\cite{Entem:2017gor} for more details.
Regarding the 3N force used with  the NV potentials, it is simply the N2LO
version with the relative LECs $c_{1,3,4}$ fixed at two-body level.
The adopted values of $c_{1,3,4}$ in the 3N force for all cases 
are reported in Table~\ref{tab:potlist}, as well.

In the last two columns of Table~\ref{tab:potlist}, we report the $\het$ and $\heq$ binding energies obtained with
the corresponding Hamiltonian. As it can be seen, the binding energies obtained with the LO500 interaction, are
rather at variance with respect to the experimental ones, $B_{\rm exp}(\het)=7.72$ MeV and $B_{\rm exp}(\heq)=28.30$ MeV. Such
binding energies are well reproduced only by including the 3N force, and fitting one of the free
parameters (in particular, $c_E$) to reproduce the tritium binding energy. Once that binding energy
is reproduced, those of $\het$ and $\heq$ turn out to be well reproduced as well. 
\begin{table*}
  \centering
  \begin{tabular}{cccccc ccccccc}
    \hline
    \hline
    Name & DOF & $O_\chi$ & $(R_{\rm S},R_{\rm L})$ or $\Lambda$ & $E$ range & Space & $c_1$ & $c_3$ & $c_4$ & $c_D$ & $c_E$ & $B(\het)$ & $B(\heq)$ \\
    \hline
    LO500       & $\pi,N$  & LO   & $500$ MeV      & 0--300 MeV & $p$ & $-$ & $-$ & $-$ &  $-$ & $-$ & $10.407$ & $40.09$ \\
    NLO500      & $\pi,N$  & NLO  & $500$ MeV      & 0--300 MeV & $p$ & $-$ & $-$ & $-$ &  $-$ & $-$ & $7.579$ & $27.47$ \\
    N2LO500/3N     & $\pi,N$  & N2LO & $500$ MeV      & 0--300 MeV & $p$ & $-0.74$ & $-3.61$ & $2.44$ & $-0.86$ & $-0.32$ & $7.727$ & $28.03$ \\
    N3LO500/3N     & $\pi,N$  & N3LO & $500$ MeV      & 0--300 MeV & $p$ & $-1.20$ & $-4.43$ & $2.67$ & $-2.351$ & $-0.723$ & $7.725 $ & $28.00$ \\
    N4LO500/3N     & $\pi,N$  & N4LO & $500$ MeV      & 0--300 MeV & $p$ & $-0.73$ & $-3.38$ & $1.69$ & $-2.234$ & $-0.384$ & $7.723$ & $28.23$ \\
    N4LO450/3N     & $\pi,N$  & N4LO & $450$ MeV      & 0--300 MeV & $p$ & $-0.73$ & $-3.38$ & $1.69$ & $-0.520$ & $+0.216$ & $7.716$ & $28.59$ \\
    N4LO550/3N     & $\pi,N$  & N4LO & $550$ MeV      & 0--300 MeV & $p$ & $-0.73$ & $-3.38$ & $1.69$ & $-2.848$ & $-0.810$ & $7.729$ & $28.13$ \\
    \hline
    NVIa/3N  & $\pi,N,\Delta$ & N3LO & $(0.8,1.2)$ fm & 0--125 MeV & $r$ & $1.2$ & $0.8$ & $1.0$ & $-0.635$ & $-0.090$ & $7.730$ & $28.24$ \\
    NVIb/3N  & $\pi,N,\Delta$ & N3LO & $(0.7,1.0)$ fm & 0--125 MeV & $r$ & $1.0$ & $0.7$ & $1.0$ & $-4.71$  & $+0.55$  & $7.727$ & $28.21$ \\
    NVIIa/3N & $\pi,N,\Delta$ & N3LO & $(0.8,1.2)$ fm & 0--200 MeV & $r$ & $1.2$ & $0.8$ & $1.0$ & $-0.61$  & $-0.35$  & $7.728$ & $28.08$ \\
    NVIIb/3N & $\pi,N,\Delta$ & N3LO & $(0.7,1.0)$ fm & 0--200 MeV & $r$ & $1.0$ & $0.7$ & $1.0$ & $-5.25$  & $+0.05$  & $7.727$ & $28.11$ \\    
    \hline
    \hline    
  \end{tabular}
  \caption{\label{tab:potlist}Summary of the properties of the NN and 3N interactions used in this
  study. In the first column we indicate the name adopted to identify each interaction
  and in the remaining columns we list its main features, including degrees of freedom (DOF), chiral order ($O_\chi$),
  cutoff values, lab-energy range over which the fits to the NN database have been carried out ($E$ range), and
  whether it is  expressed in configuration ($r$) or in momentum ($p$) space.
  All interactions, except for the first two,  include also a 3N interaction derived at N2LO.
  In columns 7-11, the values of some of the most relevant LECs are explicitly reported.
  Finally, in the last two columns, we report the $\het$ and $\heq$ binding energies
  obtained with the corresponding interaction.}
  \end{table*}

In Table~\ref{tab:currents} we report the contributions order-by-order of the axial and vector currents we use in this work. The adopted models for the nuclear axial and vector
currents are the ones derived in Refs.~\cite{Baroni:2015uza,Baroni:2018fdn} for the NV potentials
and Refs.~\cite{Pastore:2009is,Piarulli:2012bn} for the EMN ones, respectively, and are consistent 
(as possible) with the nuclear interactions used. Notice that in this work, we assume the power counting proposed by the JLab-Pisa group only (for more information
regarding this issue, see Ref.~\cite{Gnech:2023mvb}). The only independent LEC that appears up to N3LO is $d_R$ that is fixed fitting 

simultaneously the $^3$H binding energy and the GTME of the $^3$H $\beta$-decay, as already mentioned.
\begin{table*}
  \centering
  \begin{tabular}{lcccc}
    \hline
    \hline
    Oper. & LO($Q^{-3}$)  & NLO($Q^{-2}$)  & N2LO($Q^{-1}$)  & N3LO($Q^{0}$)  \\
    \hline
     $\rho(A)$    & --     & 1b(NR) & OPE & --  \\
     \hline
     ${\bm J}(A)$ & 1b(NR) & --     & 1b(RC)+OPE($\Delta$)$^*$ & OPE+CT($d_R$) \\
    \hline
      $\rho(V)$   & 1b(NR) & --     & 1b(RC) & OPE \\ 
    \hline
    ${\bm J}(V)$  & --     & 1b(NR) & OPE & 1b(RC)+OPE($\Delta$)$^*$ \\
    \hline
    \hline
  \end{tabular}
  \caption{\label{tab:currents} The power counting scheme of the weak
  axial ($A$) and vector ($V$) currents adopted in this work.
    The acronym stands for 1b=one-body, OPE=one-pion exchange,
    CT=contact terms, NR=non-relativistic, RC=relativistic corrections, OPE($\Delta$) = one-pion-exchange currents with
    an intermediate $\Delta$-isobar excitation. With the asterisk we indicate the terms that do not appear for the EMN
    interactions. The power counting for the electromagnetic current is identical
    to the counting of the weak vector current. In this table $Q$ indicates the ratio between the typical momentum of the process ($\simeq m_\pi$) and the breakdown scale of the theory ($\simeq 500$ MeV).
    }
  \end{table*}

In order to compare with the old calculation of Ref.~\cite{Marcucci:2001dmc}, we have also
employed the phenomenological AV18/UIX interaction~\cite{Wiringa:1994wb,Pudliner:1995wk}. We have used a chiral version
of the axial current where $d_R$ is fixed to reproduce the GTME for the $\tri$ $\beta$-decay, namely,
with the $\tri$ and $\het$ wave functions determined using the AV18/UIX interaction.

\subsection{The wave functions}
\label{sec:wf}

In this subsection, we report only the expression of the initial wave functions which will enter in the matrix elements
of the processes discussed later. The center-of-mass (CM) wave functions for a relative impulse $\bmp$ between the
incident particles along $z$ can be expanded in terms of components of given total angular momentum $J,J_z$ as
\begin{eqnarray}
  \Psi_{s_3,s_1}^{ph}&=& \sum_{S,L,J,J_z} \sqrt{4\pi} i^L ({1\over2},s_3,{1\over2},s_1| S, J_z)  \nonumber \\
    & \times& \!\!\!(L,0,S,J_z| J,J_z)  \sqrt{2L+1} e^{i\sigma_L} \Psi^{ph,LSJJ_z}  , \label{eq:psiph}\\
  \Psi_{s_3,s_1}^{nh}&=& \sum_{S,L,J,J_z} \sqrt{4\pi} i^L ({1\over2},s_3,{1\over2},s_1| S, J_z)  \nonumber \\
  &\times& \!\!\!(L,0,S,J_z| J,J_z)  \sqrt{2L+1}  \Psi^{nh,LSJJ_z} , \label{eq:psinh}
\end{eqnarray}
where $L$ and $S$ are the total orbital angular momentum and total spin of the incident clusters. 
Above $s_3$ ($s_1$) is the $z$-projection of the total spin of the incoming trinucleon systems (nucleon).
 $ \Psi^{ph,LSJJ_z}$ ($\Psi^{nh,LSJJ_z}$) is the $p+\het$ ($n+\het$) wave function of total
angular momentum quantum numbers $J,J_z$ and parity given by $(-)^L$. Moreover,
$\sigma_L$ is the Coulomb phase shift.
The final state   wave function of $\heq$ will be denoted using only $\Psi_4$, being the $\heq$ a state of zero total angular momentum
and even parity. The wave functions $\Psi_4$, $ \Psi^{ph,LSJJ_z}$, and $\Psi^{nh,LSJJ_z}$ are determined
using the HH method that is discussed in great detail in Refs.~\cite{Kievsky:2008es,10.3389/fphy.2020.00069,Viviani:2020gkm} and summarized for this specific application in Appendix~\ref{app:a}.

\subsection{Calculation of the hep S-factor}
\label{sec:sf}

In first-order perturbation theory, the capture matrix element $T_{fi} = \langle \heq,e^+,
\nu_e | H_{weak} | p+\het\rangle$
with the emission of a positron (neutrino) of momentum $\bmp_e$ ($\bmp_\nu$)
and spin $z$-projection $s_e$ ($s_\nu$) can be written in general as
\begin{equation}
  T_{fi}= {G_F\over \sqrt{2}} j^\mu_W(\bmq) \ell_\mu \ ,\label{eq:tfi1}
\end{equation}
where $G_F\equiv G_V\cos\theta_c\approx 1.14939\; 10^{-5}$ GeV$^2$($\theta_c$ is the Cabibbo angle),
$\ell^\mu=\overline u(\bmp_\nu,s_\nu)\gamma^\mu(1-\gamma^5) v(\bmp_e,s_e)$ is the
leptonic matrix element ($u$ and $v$ are the Dirac four-spinors), and
 $j^\mu_W(\bmq) = \langle \Psi_4 | J^\mu_W(\bmq)^\dag | \Psi_{s_3,s_1}^{ph}\rangle$ is the
matrix element of the nuclear weak current operator $J^\mu_W(\bmq)$ with $\bmq=\bmp_e+\bmp_\nu$.
The wave functions $\Psi_4$ and  $\Psi_{s_3,s_1}^{ph}$ are those introduced in Sec.~\ref{sec:wf}. 
Expressing the initial wave function in components of defined total angular momentum and parity,
as shown in Eq.~(\ref{eq:psiph}), we can perform the multipole expansion of the matrix element
(for more details, see Refs.~\cite{Walecka:1995mi,Marcucci:2001dmc}). First of all, let us write
the operator $J^\mu_W(\bmq)^\dag\equiv\{\rho_W(\bmq)^\dag,{\bm{J}_W}({\bm{q}})^{\dag}\}$.
Then, the three-dimensional part ${\bm{J}_W}({\bm{q}})^{\dag}$ can be written as
\begin{equation}
  {\bm{J}_W}({\bm{q}})^{\dag}=\{ {\hat{\bm{e}}}^{*}_{z}\cdot{\bm{J}_W}({\bm{q}})^{\dag},
  {\hat{\bm{e}}}^{*}_{\lambda}\cdot{\bm{J}_W}({\bm{q}})^{\dag} \}\ , \lambda=\pm1\ .
\end{equation}
In the above, we have introduced the basis of unit vectors
\begin{equation}
\hat{\bm e}_{z}=\hat\bmq\ ,\qquad \hat{\bm e}_{y}=\frac{{\bm p}\times\bmq}{|{\bm p}\times\bmq|}
\ ,\qquad \hat{\bm e}_{x}=\hat{\bm e}_{y}\times\hat{\bm e}_{z} \ ,
\end{equation}
and ${\bm e}_\pm=\mp (\hat{\bm e}_x\pm i\, \hat{\bm e}_y)/\sqrt{2}$.
The matrix elements of these components can then be written as
\begin{eqnarray}
 \lefteqn{    \langle\Psi_{4}|\rho^{\dag}({\bm{q}})|\Psi^{ph,LSJJ_{z}}\rangle 
   =\qquad\qquad} &&  \nonumber \\
  &&  \sqrt{4\pi} (-{\rm{i}})^{J} (-)^{J-J_z} D_{-J_{z},0}^{J}(-\phi,-\theta,0) C_{J}^{LSJ}(q) \ , \label{eq:c}\\
   \lefteqn{  \langle\Psi_{4}|{\hat{\bm{e}}}^{*}_{z}\cdot
   {\bm{J}_W}^{\dag}({\bm{q}})|\Psi^{ph,LSJJ_{z}}\rangle
   =\qquad\qquad} &&  \nonumber \\
  &&   \sqrt{4\pi} (-{\rm{i}})^{J}  (-)^{J-J_z}
   D_{-J_{z},0}^{J}(-\phi,-\theta,0) L_{J}^{LSJ}(q) \ , \label{eq:l} \\
 \lefteqn{    \langle\Psi_{4}|{\hat{\bm{e}}}^{*}_{\lambda}\cdot
   {\bm{J}_W}^{\dag}({\bm{q}})|\Psi^{ph,LSJJ_{z}}\rangle
   =\qquad\qquad} &&  \nonumber \\
    && -\sqrt{2\pi} (-{\rm{i}})^{J} (-)^{J-J_z}  D_{-J_{z},-\lambda}^{J}(-\phi,-\theta,0) \nonumber\\
    &&\times \left[ \lambda M_{J}^{LSJ}(q) +E_{J}^{LSJ}(q)\right] \ . \label{eq:me}
\end{eqnarray}
where $C_{\ell}^{LSJ}$, $E_{\ell}^{LSJ}$, $M_{\ell}^{LSJ}$, and $L^{LSJ}_\ell$ denote the reduced matrix 
elements (RMEs) of the charge $(C)$, transverse electric $(E)$,
transverse magnetic $(M)$, and longitudinal $(L)$ multipole operators~\cite{Walecka:1995mi,Marcucci:2001dmc}. The weak charge/current operators  have components of both scalar/polar-vector ($V$) 
and pseudoscalar/axial-vector ($A$) character, hence 
\begin{equation}
T^{LSJ}_{J}(q)=T^{LSJ}_{J}({V})+T^{LSJ}_{J}({A}) \ ,
\label{eq:tav}
\end{equation}
where $T^{LSJ}_{J}$ is any of the RMEs introduced above (in this notation, the dependence on $q$ is understood).
Since the initial/final wave functions have a definite parity,
only one between $T^{LSJ}_{J}({V})$ and $T^{LSJ}_{J}({A})$ is different from zero (see below).
In the above expressions, the spin quantization axis of the nuclear states is taken along
the incident relative momentum ${\bm p}\,$=$\,p\, \hat{\bm z}$ rather
than the three-momentum transfer $\bmq\,$=$\,q\, \hat{\bm e}_z$.
For this reason, we needed to introduce the Wigner rotation 
matrices ${D}_{M',M}^{J}$~\cite{Edmond1957}. The angles
$\theta$ and $\phi$ specify the direction of $\bmq$
in the lab frame (with ${\bm p}$ along $\hat{\bm z}$).
In any case, the RMEs can be computed in a frame where $\bmq$ is along ${\hat{\bm z}}$, where
$D^J_{M',M}(0,0,0)=\delta_{M',M}$.

The cross section for the $^{3}$He($p$,$e^{+}\nu_{e}$)$^{4}$He reaction 
at a CM\ energy $E$ is given by
\begin{eqnarray}
  \sigma(E)&=& \int 2\pi \, \delta\left (\Delta m  + E -  \frac{q^{2}}{2 M_{4}} - E_e- E_\nu\right )\frac{1}{v_{\rm rel}} \nonumber \\
     &&\times \frac{1}{4}\sum_{s_e s_\nu}\sum_{s_1 s_3} 
    |T_{fi}|^{2} 
    \frac{d{\bm{p}}_{e}}{(2\pi)^3} \frac{d{\bm{p}}_{\nu}}{(2\pi)^3} \ ,
   \label{eq:xsc1}
\end{eqnarray}
where $\Delta m = M_p + M_3 - M_4 $ = 19.287 MeV
with $M_p$, $M_3$, and $M_4$ being the proton, $\het$, and $\heq$ rest masses, respectively, and $v_{\rm rel}$
is the $p\,^{3}$He relative velocity. 
Moreover, $E_e=\sqrt{p_e^2+m_e^2}$ and $E_\nu=p_\nu$ ($m_e$ is the positron
mass and we can safely neglect the neutrino mass). 
The energy $E$ and the relative velocity $v_{\rm rel}$ are related
to the CM relative momentum $\bmp$ as $E=p^2/2\mu$ and $v_{\rm rel }=p/\mu$,
$\mu$ being the $p\het$ reduced mass.
From Eq.~(\ref{eq:tfi1}) we have
\begin{equation}
   \frac{1}{4}\sum_{s_e s_\nu}\sum_{s_1 s_3} 
   |T_{fi}|^{2} =  {G_F^2\over 2}\> L_{\sigma \tau} \>
   N^{\sigma\tau} ,
\label{eq:lwst}
\end{equation}
where the lepton tensor $L^{\sigma \tau}$ is defined as
\begin{eqnarray}
  L^{\sigma \tau}&\equiv&  \sum_{s_e s_\nu}\ell^{\sigma}{\ell^{\tau}}^{*} \nonumber \\
  &=&{\rm tr}
  \bigg[\gamma^{\sigma}(1-\gamma_5)\frac{({\not{p}}_{e}-m_{e})}{2E_{e}}
  \gamma^{\tau}(1-\gamma_5)\frac{{\not{p}}_{\nu}}{2E_{\nu}}\bigg] \nonumber \\
  &=&2[v_{e}^{\sigma}v_{\nu}^{\tau}+
  v_{\nu}^{\sigma}v_{e}^{\tau}-g^{\sigma\tau}
  v_{e} \cdot v_{\nu}\nonumber \\
  &&+
  {\rm{i}}\>\epsilon^{\sigma\alpha\tau\beta}
  v_{e,\alpha} v_{\nu,\beta}]\ , 
\label{eq:lalb}
\end{eqnarray}
with $\epsilon^{0123}=-1$,  $v_e^\sigma=p_e^\sigma/E_e$ and
$v_\nu^\sigma=p_\nu^\sigma/E_\nu$. 
The nuclear tensor $N^{\sigma \tau}$ is defined as
\begin{equation}
  N^{\sigma \tau} \equiv {1\over 4}\sum_{s_1 s_3} j_W^\sigma({\bm q})
  j_W^{\tau *}({\bm q}) \ ,
  \label{nuclt}
\end{equation}
where we rewrite the quantities $j_W^\sigma({\bm q})=  \langle \Psi_4 | J^\mu_{\rm W}(\bmq)^\dag | \Psi_{s_3,s_1}^{ph}\rangle$ in terms of RMEs using 
the expansion of Eq.~(\ref{eq:psiph}) of the initial wave function and the expressions
given in Eqs.~(\ref{eq:c})--(\ref{eq:me}).

The energy-conservation delta function in Eq.~(\ref{eq:xsc1}) reduces the integration to five variables,
the spherical angles of the positron and neutrino, plus the positron
momentum. This five-fold integration is performed numerically. In such integration, one
needs the value of the RMEs calculated for many values of $q=|\bmp_e+\bmp_\nu|$.
The RMEs are tabulated in a grid of $q$ values up to the maximum value $q_{max}\approx 20$ MeV/c, and
interpolated at the specific values of $q$ needed in the integral for the cross section. Since $q_{max}$ is small, the $q$ dependence
of the RMEs is simple (actually, they can be well reproduced by powers of $q$), so this
part does not present any criticality. We also obtain the differential cross section $d\sigma/dE_e$ by writing $dp_e=(E_e/p_e)dE_e$, and then not performing the integration over $E_e$.

Once the total cross section has been obtained, the S-factor can be easily calculated as
\begin{equation}
S(E) = E\, \sigma(E)\,
{\rm exp}( 4\, \pi \, \alpha/v_{\rm rel}) \ ,
\end{equation}
where $\sigma(E)$ is the hep cross section of Eq.~\eqref{eq:sigma}, $\alpha$ the fine structure constant, and $v_{\rm rel}$ the $p-\het$ relative velocity in the CM reference frame.

\subsection{Calculation of the hen cross section}
\label{sec:xs}

For the hen process, the first-order perturbation
theory  matrix element $T^{\rm EM}_{fi} = \langle \heq,\gamma | H_{\rm EM} | n+\het\rangle$ involves
the electromagnetic (EM) interaction only. This can be written  as
\begin{equation}
  T_{fi}^{\rm EM}= {e\over\sqrt{2q}} j^\mu_{\rm EM}(\bmq) \epsilon_\mu^*({\bmq,\lambda})\ ,\label{eq:tfi2}
\end{equation}
where $e$ is the unit charge ($e^2=4\pi\alpha$), $\bmq$ the photon momentum, $\epsilon^\mu({\bmq,\lambda})$ its
polarization four-vector with $\lambda=\pm1$.
Now, $j^\mu_{\rm EM}(\bmq) = \langle \Psi_4 | J^\mu_{\rm EM}(\bmq)^\dag | \Psi_{s_3,s_1}^{nh}\rangle$ is the
matrix element of the nuclear EM current operator $J^\mu_{\rm EM}(\bmq)$ and the wave function 
$\Psi^{nh}_{s_3,s_1}$ is given in Eq.~(\ref{eq:psinh}).
As before, the matrix element $j^\mu_{\rm EM}(\bmq)$ can be decomposed in terms of multipoles with the difference that only the magnetic and electric RMEs enter in the final expression
due to the transverse polarization of the final real photon. 
 The cross section reads 
\begin{eqnarray}
  \sigma(E)&=& \int 2\pi \, \delta\left (\Delta m  + E -  \frac{q^{2}}{2 M_{4}} - q\right )\frac{1}{v_{\rm rel}} \nonumber \\
     &&\times \frac{1}{4}\sum_{\lambda}\sum_{s_1 s_3} 
    |T_{fi}^{\rm EM}|^{2} 
    \frac{d{\bm{q}}}{(2\pi)^3} \ ,
   \label{eq:xsc2}
\end{eqnarray}
where $\Delta m = M_n + M_3 - M_4 $ = 20.577 MeV with $M_n$ the neutron mass, and $v_{\rm rel}$
 the $n\,^{3}$He relative velocity. In this case
the integrations can be performed analytically and the $\het(n,\gamma)\heq$ radiative
cross section can be directly expressed in terms of RMEs as
\begin{eqnarray}
  \sigma(E) &=& {8 \pi^2\alpha\over v_{\rm rel}}  {q\over 1+q/M_4} \!\!\sum_{LS,J\geq1}
  \Bigl[ |E_J^{LSJ}({\rm EM})|^2 \nonumber \\
  && \qquad\qquad \qquad+ |M_J^{LSJ}({\rm EM})|^2\Bigr] ,\label{eq:capture}
\end{eqnarray}
with the value of $q$ fixed by the delta energy conservation function and the various 
RMEs denoted as $T^{LSJ}_J({\rm EM)}$, $T\equiv E,M$ for the electric and magnetic multipoles, respectively.

\subsection{Estimation of the theoretical uncertainty}
\label{sec:teoun}

The analysis of the uncertainty due to the truncation of the chiral expansion
has been performed using the  method proposed in Ref.~\cite{Epelbaum:2014efa}.
We choose this approach for two reasons: (i) a full Bayesian analysis such as the one in Ref.~\cite{Melendez:2019izc} will require a larger set of calculations; (ii) the method is equivalent 
to choose uniform priors in the Bayesian analysis and, as shown in Ref.~\cite{Gnech:2023mvb}, it gives results similar as compared to a more complete and sophisticated Bayesian approach. 
This has been confirmed for the proton-proton fusion reaction in Ref.~\cite{Barlucchi:2026}.
In this section we briefly discuss  this procedure that we have applied to the calculations performed using
the EMN family of interactions and accompanying currents.

First of all, we define the ``expansion parameter'' of the
chiral effective theory as $x=Q/\Lambda_b$ where $Q$ is a typical value of the nucleon momenta in nuclei  and $\Lambda_b$ the %breaking scale energy 
breakdown energy scale
of our theory. Namely, $Q$ is the low-energy scale and $\Lambda_b$ the high-energy scale of the physics
not resolved by the effective theory and taken into account via the LECs. In the following we assume $Q=m_\pi$, where $m_\pi$ is the pion mass, and 
 $\Lambda_b=500$ MeV the
 breakdown scale of our theory, for which we have $x\approx 0.3$.

We identify two sources of uncertainties that are treated separately.
\begin{itemize}
  \item Uncertainty associated to the truncation of the chiral expansion of the interaction. Following Ref.~\cite{Epelbaum:2014efa} we define the quantities
    \begin{eqnarray}
  \Delta_{\text{NLO}}({\cal O})  &=& |{\cal O}_{\text{NLO}}-{\cal O}_{\text{LO}}| \ , \label{eq:dev1}\\
  \Delta_{\text{N2LO}}({\cal O})  &=&  |{\cal O}_{\text{N2LO}}-{\cal O}_{\text{NLO}}| \ ,\label{eq:dev2} \\
  \Delta_{\text{N3LO}}({\cal O})  &=&  |{\cal O}_{\text{N3LO}}-{\cal O}_{\text{N2LO}}| \ , \label{eq:dev3}\\
  \Delta_{\text{N4LO}}({\cal O})  &=&  |{\cal O}_{\text{N4LO}}-{\cal O}_{\text{N3LO}}| \ .\label{eq:dev4}
\end{eqnarray}
with ${\cal O}_{\text{LO}}$, $\ldots$, ${\cal O}_{\text{N4LO}}$ the observable
    calculated with the interaction truncated at order LO, $\ldots$, N4LO, respectively. Using the prescription of
Ref.~\cite{Epelbaum:2014efa}, we use the following expressions for
    the uncertainty order-by-order 
\begin{eqnarray}
  \delta{\cal O}^I_{\text{NLO}}&=& \max\Bigl[x^3 {\cal O}_{\text{LO}}, x\Delta_{\text{NLO}}({\cal O})\Bigr]\ ,\label{eq:unv1}\\
  \delta{\cal O}^I_{\text{N2LO}}&=& \max\Bigl[x^4 {\cal O}_{\text{LO}}, x^2\Delta_{\text{NLO}}({\cal O}),  \nonumber\\
    && \qquad x \Delta_{\text{N2LO}}({\cal O}) \Bigr]\ ,\label{eq:unv2}\\
 \delta{\cal O}^I_{\text{N3LO}}&=& \max\Bigl[x^5 {\cal O}_{\text{LO}}, x^3\Delta_{\text{NLO}}({\cal O}), 
   x^2 \Delta_{\text{N2LO}}({\cal O}),\nonumber\\
   && \qquad x \Delta_{\text{N3LO}}({\cal O}) \Bigr]\ ,\label{eq:unv3}\\
 \delta{\cal O}^I_{\text{N4LO}}&=& \max\Bigl[x^6 {\cal O}_{\text{LO}}, x^4\Delta_{\text{NLO}}({\cal O}), 
   x^3 \Delta_{\text{N2LO}}({\cal O}), \nonumber\\
   && \qquad x^2 \Delta_{\text{N3LO}}({\cal O}), 
 x \Delta_{\text{N4LO}}({\cal O}) \Bigr]\ .\label{eq:unv4}
\end{eqnarray}
  Passing from the LO to NLO, $x$ jumps by two powers instead of one because
    the NLO interaction includes terms of order $Q^0$ and $Q^2$, being  the terms $\propto Q^1$ identically zero. All the other orders gain only one power of $Q$.
\item Uncertainty due to the truncation of the chiral expansion of the current. 
    Now ${\cal O}_{\text{LO}}$, $\ldots$, ${\cal O}_{\text{N3LO}}$ are the values of the observable ${\cal O}$
    calculated with the current truncated at order LO, $\ldots$, N3LO, respectively (but for a fixed order
    of the interaction). We consider here the current terms up to N3LO and in this case
    there are no ``jumps''. As before, we compute the quantities $\Delta$ as in Eqs.~(\ref{eq:dev1})--(\ref{eq:dev3}).
    Then, the method of Ref.~\cite{Epelbaum:2014efa} is applied to the current expansion as:
\begin{eqnarray}
  \delta{\cal O}^C_{\text{NLO}}&=& \max\Bigl[x^2 {\cal O}_{\text{LO}}, x\Delta_{\text{NLO}}({\cal O})\Bigr]\ ,\label{eq:unj1}\\
  \delta{\cal O}^C_{\text{N2LO}}&=& \max\Bigl[x^3 {\cal O}_{\text{LO}}, x^2\Delta_{\text{NLO}}({\cal O}),  \nonumber\\
    && \qquad x \Delta_{\text{N2LO}}({\cal O}) \Bigr]\ ,\label{eq:unj2}\\
 \delta{\cal O}^C_{\text{N3LO}}&=& \max\Bigl[x^4 {\cal O}_{\text{LO}}, x^3\Delta_{\text{NLO}}({\cal O}), 
   x^2 \Delta_{\text{N2LO}}({\cal O}),\nonumber\\
   && \qquad x \Delta_{\text{N3LO}}({\cal O}) \Bigr]\ .\label{eq:unj3}
\end{eqnarray}
\end{itemize}

To give a more statistical insight,  we assume that the value of the observable $\cal O$ is
 uniformly distributed within the $\pm\delta{\cal O}^{I/C}_{\rm NiLO}$ and
 so the $68\%$ confidence level (CL) is given by the value of the truncation error divided
 by $\sqrt{3}$, that at order $i$ reads
\begin{equation}\label{eq:sigma}
    \sigma^{I/C}_{i}=\frac{\delta {\cal O}^{I/C}_{\text{N}i\text{LO}}}{\sqrt{3}}
\end{equation}
To combine the results obtained using different orders of the interactions and currents,
we follow Ref.~\cite{Gnech:2023mvb}. Following this approach,
we assume that the truncation uncertainties are fully correlated  obtaining the most conservative estimate by summing the single truncation uncertainties, i.e.
\begin{equation}
    \sigma_{\chi {\rm EFT}}=\sigma^{I}_{i}+\sigma^{C}_{j}\,,
\end{equation}
where $i$ and $j$ are the order of the interaction and current respectively.

The model averaging on the $k$ interaction models is then obtained using
\begin{equation}\label{eq:meanO}
\langle {\cal O} \rangle =\sum_k {\cal O}_k\; {\rm pr}(k)\,,
\end{equation}
for the central value and 
\begin{equation}\label{eq:systO}
\sigma^2_{\cal O}=\sum_k \sigma^2_{\chi {\rm EFT},k}\; {\rm pr}(k)+\sigma^2_{\cal O,{\rm syst}}\,,
\end{equation}
for the variance. In the following, we indicate with $k=1,2,3$
the models N4LO450/3N, N4LO500/3N, and N4LO550/3N, and with $k=4,5,6,7$ the models
NVIa/3N, NVIb/3N, NVIIa/3N, NVIIb/3N.
The systematic error  $\sigma^2_{\cal O,{\rm syst}}$ represents the model dependence and is computed as
\begin{equation}
\sigma^2_{\cal O,{\rm syst}}=\sum_k {\cal O}_k^2 \; {\rm pr}(k)-\left(\sum_k {\cal O}_k\; {\rm pr}(k)\right)^2\,.
\end{equation}
In these equations,  "${\rm pr}$" is given by
\begin{equation}
    {\rm pr}(k)=\begin{cases} \frac{1}{8}, & \mbox{if } k\in\mbox{ NV family} \\ \frac{1}{6}, & \mbox{if } k\in\mbox{ EMN family} \end{cases}\,.
\end{equation}
This choice is made in order not to privilege neither the EMN nor the NV models.

\section{Results}
\label{sec:res}
In this section, we present our results. The section is organized as follows: in Sec.~\ref{subsec:hep-elastic} we test the quality of the $p+\het$ scattering wave function comparing our predictions for the elastic cross section and polarization observables with the available experimental data at a given CM energy. In Sect.~\ref{subsec:hep0rme} we consider the hep process at zero CM energy and discuss the results for the RMEs calculated in the various scattering channels. In Sects.~\ref{subsec:hep0S} and~\ref{subsec:hep100} we present the astrophysical S-factor calculated at zero CM energy and up to 100 keV. In this way, we will be able to extract, in addition to $S(0)$, also the first derivative for $S(E)$. Finally in Secs.~\ref{subsec:hen0} and~\ref{subsec:hen1} we present the hen cross section calculated at thermal energies or up to a few MeV in the CM frame, respectively.

\subsection{$p+\het$ elastic scattering}
\label{subsec:hep-elastic}
To prove the quality of our wave functions and the adopted nuclear interactions, we present the results obtained for various observables for $p+\het$ elastic scattering at $E=4.15$ MeV, for which accurate experimental data are available~\cite{McDonald:1964zz,Alley:1993zza,Daniels:2010af}.

\begin{figure}[bth]
\centering
\includegraphics[scale=0.40,clip]{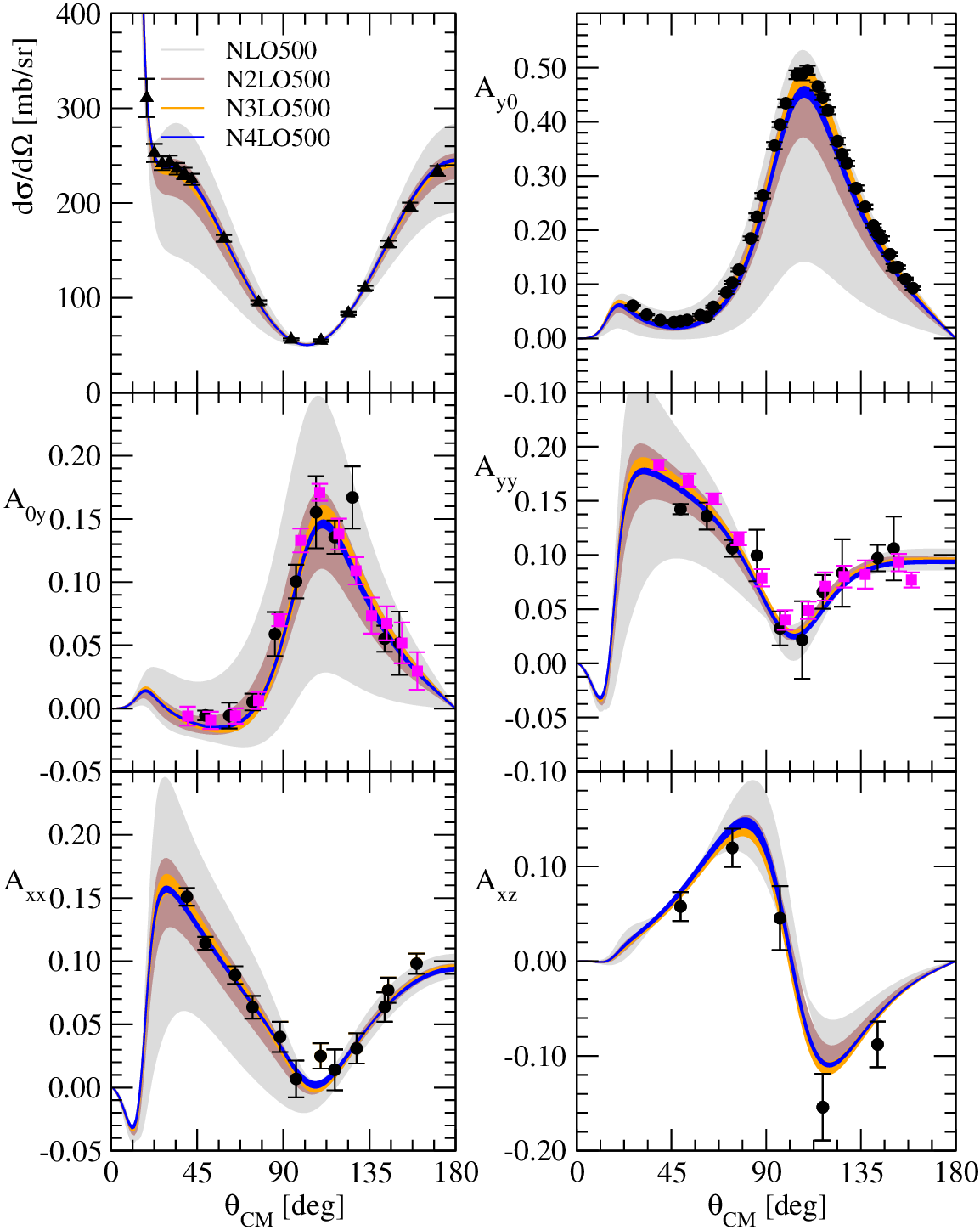}
\caption{(color online) Calculated observables for $p+\het$ elastic scattering at $E=4.15$ MeV,
   using the EMN interactions with fixed cutoff $\Lambda=500$ MeV, from LO to N4LO.
   The width of each band is the theoretical uncertainty, calculated as discussed in the main text.
   The grey, brown, orange, and blue bands correspond to the calculations performed at NLO, N2LO, N3LO,
   and N4LO chiral orders, respectively.
   The experimental values are taken from Ref.~\protect\cite{McDonald:1964zz}
   (solid black upper triangles), 
  Ref.~\protect\cite{Alley:1993zza} (solid black circles),
  and Ref.~\protect\cite{Daniels:2010af} (solid magenta squares).}
  \label{fig:ph1}
\end{figure}

In Fig.~\ref{fig:ph1}, we report the results of the calculations obtained using the EMN interactions, from LO500 up to N4LO500. In the figure, we plot the $3\sigma$ error band where the variance is obtained in Eq.~\eqref{eq:sigma} from  the uncertainty computed using the method introduced in Ref.~\cite{Epelbaum:2014efa} and briefly recalled in Sec.~\ref{sec:teoun}. 
As can be seen, the error bands  show a good convergence pattern 
for all observables.

From the figure it is also clear that the combination of the HH scattering wave function with the EMN interaction with $\Lambda=500$ MeV gives a good description of the differential cross section and the analyzing powers. Similar results can be found for the NV family of interactions. Therefore, we conclude that the EMN and NV families of interactions well describe the nuclear dynamics of $A=4$ systems. 
However, it is important to note that, from the quantitative point of view, some discrepancy still remains. As an example, in Table~\ref{tab:ay0} we report the error analysis of the 
 analyzing power $A_{y0}$ at  maximum angle, i.e. at $\theta_{CM}=110$ degrees. 
 \begin{table}[]
     \centering
     \begin{tabular}{lcccc}
     \hline\hline
         $i$ & $A_{y0}$ & $\Delta(A_{y0})$ & $\delta A_{y0}$ & $3\times\sigma^I_i$ \\
         \hline
         LO & -0.0311 & & &  \\
         NLO & 0.3343 & 0.3654 & 0.1008 & 0.1745 \\

         N2LO & 0.4225 & 0.0882 & 0.0278 & 0.0482 \\
         N3LO & 0.4721 & 0.0496 & 0.0138 & 0.0239\\
         N4LO & 0.4530 & 0.0191 & 0.0052 & 0.0090 \\
         \hline\hline
     \end{tabular}
     \caption{Error analysis of the $p+\het$ elastic scattering analyzing power $A_{y0}$ at $\theta_{CM}=110$ deg corresponding to the maximum value of the observable. The last column correspond to the $3\sigma$ uncertainty. The cutoff $\Lambda$ is fixed at 500 MeV. }
     \label{tab:ay0}
 \end{table}
 The calculated value for the N4LO500/3N interaction,
considering the $3\sigma$ uncertainty, is
 $A_{y0}(\theta_{CM}=110\ {\text{deg}})=0.4530\pm0.0090$, to be compared with
the experimental value $0.4952\pm 0.0082$ \cite{Alley:1993zza}. The small discrepancy is thought to be due to missing high-order contributions in the 3N force.

\subsection{hep at zero energy: RMEs}
\label{subsec:hep0rme}
In this section, we consider the hep process at zero energy, and in particular the values for the multipole RMEs.
In Table~\ref{tab:rmes} we list the most relevant multipoles contributing to the transition
from an initial $^{2S+1}L_J$ $p+\het$ scattering state to the final $J^\pi\,$=$0^+$ $\heq$ ground state.
Note that the multipolarity $\ell$ of the RME
is fixed to $\ell=J$. Moreover, the parity condition selects a given RME as
coming from either the vector or the axial part of the weak current only. 

\begin{table}[bth]
    \begin{center}
      \begin{tabular}{l|c|cccc}
        \hline\hline
        $J^\pi$ & ${}^{2S+1}L_J$ & $C_\ell^{LSJ}$ & $E_\ell^{LSJ}$ & $M_\ell^{LSJ}$ & $L_\ell^{LSJ}$ \\
        \hline
        $0^+$  &  ${}^1S_0$ & $C_0^{000}(V)$ & $L_0^{000}(V)$ & $-$ & $-$ \\
        $0^-$  &  ${}^3P_0$ & $C_0^{110}(A)$ & $L_0^{110}(A)$ & $-$ & $-$ \\
        $1^+$  &  ${}^3S_1, {}^3D_1$ & $C_1^{LS1}(A)$ & $L_1^{LS1}(A)$ & $E_1^{LS1}(A)$ & $M_1^{LS1}(V)$ \\
        $1^-$  &  ${}^1P_1, {}^3P_1$ & $C_1^{LS1}(V)$ & $L_1^{LS1}(V)$ & $E_1^{LS1}(V)$ & $M_1^{LS1}(A)$ \\
        $2^+$  &  ${}^1D_2, {}^3D_2$ & $C_2^{LS2}(V)$ & $L_2^{LS2}(V)$ & $E_2^{LS2}(V)$ & $M_2^{LS2}(A)$ \\
        $2^-$  &  ${}^3P_2, {}^3F_2$ & $C_2^{LS2}(A)$ & $L_2^{LS2}(A)$ & $E_2^{LS2}(A)$ & $M_2^{LS2}(V)$ \\
        $3^+$  &  ${}^3D_3, {}^3G_3$ & $C_3^{LS3}(A)$ & $L_3^{LS3}(A)$ & $E_3^{LS3}(A)$ & $M_3^{LS3}(V)$ \\
        \hline
        \hline
      \end{tabular}    
 \caption{\label{tab:rmes}
   The RMEs $C^{LSJ}_\ell$, $E^{LSJ}_\ell$, $M^{LSJ}_\ell$, and $L^{LSJ}_\ell$ contributing
   to the hep process from an initial $3+1$ $^{2S+1}L_J$ scattering state to the final
   $\heq$  ground state. Note that either the vector $(V)$ or the axial $(A)$ part of the weak current can contribute.}
   \end{center}
  \end{table}

The values for the multipole
RMEs calculated at zero energy and for a lepton momentum transfer of $q$=$19.2$ MeV/c,
connecting any of the $p+\het$ S- and P-wave channels to the $^4$He bound state are shown in Tables~\ref{tab:hep1},~\ref{tab:hep2}, and~\ref{tab:hep3}.
Note that the RMEs listed in all these tables are related to those defined in Eqs.~(\ref{eq:c})--(\ref{eq:me}) via 
\begin{equation}
    \overline{T_J}^{LSJ}=
      \sqrt{ {v_{\rm rel} \over 4\pi\alpha }
      [{\rm exp}(4\pi\alpha/v_{\rm rel})-1]} \,
      T_J^{LSJ} \label{newRME2} \ ,
\end{equation}
such that they remain finite in the limit $v_{\rm rel} \rightarrow 0$,
corresponding to zero energy. Moreover, we refer to Table~\ref{tab:currents}
for the various terms included in the weak transition operator.

In Table~\ref{tab:hep1} we present the convergence of the RMEs as the basis sets of HH functions
used to describe the bound and scattering states are enlarged. This convergence is shown
for the ${}^3S_1$ scattering state, which gives one of the most important contributions
to the hep S-factor, but analogous results are obtained for other waves. We consider the calculations
performed with the N4LO500/3N interaction and accompanying N3LO weak current operators.  The basis sets B (S) are those relative to the
expansion of the bound (scattering) state in HH functions (for more details see Appendix~\ref{app:a}).
In the basis set B1 (B2), the bound-state wave function is expanded in 2700 (3379) HH functions.
In the basis sets S1, $\ldots$, S4, the core part of the scattering-state wave function
is expanded in  3391, 4233, 4873, and 5609 HH functions, respectively.
As can be seen from the table, the RMEs are almost insensitive to the chosen basis.
This behavior has also been observed for other scattering waves and for all interactions
and currents considered in this work. Therefore, such basis are sufficient to achieve convergence for the binding energy or the phase shifts at the fourth digit and a specific choice is
not critical for this reaction.

\begin{table}[t]
\begin{center}
\begin{tabular}{cc|cccc}
\hline 
\hline
 Basis set B & Basis set S  & $\overline C_1(A)$  &  $\overline L_1(A)$ & $\overline M_1(V)$  &  $\overline E_1(A)$ \\
\hline
B1 & S1 & $0.01034$  & $-0.02677$ & $0.00014$ & $ -0.03986$ \\
B1 & S2 & $0.01036$  & $-0.02681$ & $0.00014$ & $ -0.03991$ \\
B1 & S3 & $0.01036$  & $-0.02683$ & $0.00014$ & $ -0.03993$ \\
B1 & S4 & $0.01037$  & $-0.02682$ & $0.00014$ & $ -0.03994$ \\
\hline
B2 & S4 & $0.01037$  & $-0.02682$ & $0.00014$ & $ -0.03994$ \\
\hline
        \hline
\end{tabular}
  \caption{\label{tab:hep1}
    RMEs $\overline C_1(A)$, $\overline L_1(A)$, $\overline M_1(V)$, and $\overline E_1(A)$ for the
    ${}^3S_1$ capture at zero $p+\het$ CM energy, calculated for various basis sets employed in the
    expansion of the bound and scattering wave functions (as specified in the first two columns).
    See the main text for more details. The calculations is performed for the
    N4LO500/3N interaction and accompanying N3LO weak current. The momentum transfer $q$ is $19.2$ MeV/c.    
    The RMEs are purely imaginary and in fm$^{3/2}$ units.}
    \end{center}
\end{table}

\begin{table}[t]
\begin{center}
\begin{tabular}{lcc}
\hline 
\hline
 & \multicolumn{2}{c}{${}^1S_0$ RMEs}\\
 Order & $\overline C^{000}_0(V)$  &  $\overline L^{000}_0(V)$ \\
\hline
  $Q^{-3}$ & $ -0.00572$ & $  0.00000 $ \\
  $Q^{-2}$ & $ -0.00572$ & $ -0.00573 $ \\
  $Q^{-1}$ & $ -0.00572$ & $ -0.00573 $ \\
  $Q^{0}$  & $ -0.00572$ & $ -0.00573 $ \\
\hline
& \multicolumn{2}{c}{${}^3P_0$ RMEs}\\
Order & $\overline C^{110}_0(A)$  &  $\overline L^{110}_0(A)$ \\
\hline
  $Q^{-3}$ & $  0.00000 $ & $  0.01905 $ \\
  $Q^{-2}$ & $  0.04082 $ & $  0.01905 $ \\
  $Q^{-1}$ & $  0.04813 $ & $  0.01917 $ \\
$Q^{0}$  & $  0.04813 $ & $  0.01902 $ \\
        \hline
        \hline
\end{tabular}
  \caption{\label{tab:hep2}
    Cumulative contributions to the ${}^1S_0$ and ${}^3P_0$ RMEs for the hep process at zero $p+\het$ CM energy.
    The momentum transfer $q$ is $19.2$ MeV/c, and the results have been obtained with
    the N4LO500/3N interaction and accompanying currents. The RMEs have been given
    for the various chiral orders of the weak current as specified in Table~\ref{tab:currents}. 
    Note that the ${}^1S_0$ (${}^3P_0$) RMEs are purely real (imaginary) and in fm$^{3/2}$ units.}\end{center}
\end{table}

In Tables~\ref{tab:hep2} we present the $J=0$  RMEs varying the order of the current, while keeping fixed the interaction N4LO500/3N.
For channel $^1S_0$, the $C_0({V})$ and $L_0({V})$ RMEs receive contributions from the vector charge and the longitudinal component of the vector current, respectively.  The associated RMEs, while small, are not negligible--they
are about 20 \% of the main contribution originating from the $E_1({A})$ RME in $^3S_1$ capture as shown
in Table~\ref{tab:hep3}.  These $^1S_0$ transitions are inhibited 
by  isospin selection rule, vanishing then at $q$=$0$,
as discussed in detail in Ref.~\cite{Marcucci:2001dmc}.
Moreover, being the weak vector current conserved, if the initial and final HH
wave functions were to be exact eigenfunctions of the Hamiltonian,
then one would expect, neglecting the kinetic energy of the recoiling $^4$He:
\begin{equation}
L_J({V}) = \frac{E_3-E_4}{q}\, C_J({V}) \ ,
\label{eq:c0l0}
\end{equation}
where $E_3$ and $E_4$ are the ground-state energies of three and four-nucleons
. For $q$=$19.2$ MeV/c the ratio $L_0/C_0$ is then
expected to be $\approx 1$, which is in agreement with that obtained
in the calculation, when the two-body current contributions are taken
into account.

The $^3P_0$ capture is induced by the weak axial charge and 
the longitudinal component of the weak axial current 
via the $C_{0}({A})$ and $L_{0}({A})$ multipoles, 
respectively. The two-body axial charge operators
at order $Q^{-1}$ give a $\simeq 20$ \% correction to the one-body
contribution of the $C_0({A})$ RME. The $L_0({A})$ RME
is dominated by the order $Q^{-3}$ axial current operator (the Gamow-Teller operator),
but its value is only half of $C_0({A})$.
The $C_0({A})$ and $L_0({A})$
RMEs are expected to have the same sign, as discussed in
Sect. VI.C of Ref.~\cite{Marcucci:2001dmc}.
This positive relative sign produces a destructive
interference between these RMEs in the cross section, substantially
reducing the $^3P_0$ overall contribution to the S-factor.

\begin{table}[t]
\begin{center}
\begin{tabular}{lcccc}
\hline 
\hline
 & \multicolumn{4}{c}{${}^3S_1$ RMEs}\\
Order & $\overline C_1(A)$  &  $\overline L_1(A)$ & $\overline M_1(V)$  &  $\overline E_1(A)$ \\
\hline
 $Q^{-3}$ & $  0.00000$ & $ -0.02686$ & $  0.00000$ & $ -0.03918$ \\
 $Q^{-2}$ & $  0.00996$ & $ -0.02686$ & $  0.00153$ & $ -0.03918$ \\
 $Q^{-1}$ & $  0.01053$ & $ -0.03464$ & $ -0.00046$ & $ -0.05072$ \\
 $Q^{0}$  & $  0.01053$ & $ -0.02682$ & $  0.00014$ & $ -0.03994$ \\
\hline
 & \multicolumn{4}{c}{${}^1P_1$ RMEs}\\
Order & $\overline C_1(V)$  &  $\overline L_1(V)$ & $\overline M_1(A)$  &  $\overline E_1(V)$ \\
\hline
 $Q^{-3}$ & $ -0.02427$ & $  0.00000$ & $ -0.00208$ & $  0.00000$ \\
 $Q^{-2}$ & $ -0.02427$ & $ -0.02432$ & $ -0.00208$ & $ -0.02823$ \\
 $Q^{-1}$ & $ -0.02439$ & $ -0.02443$ & $ -0.00187$ & $ -0.03404$ \\
 $Q^{0}$  & $ -0.02440$ & $ -0.02444$ & $ -0.00129$ & $ -0.03388$ \\
\hline
 & \multicolumn{4}{c}{${}^3P_1$ RMEs}\\
Order & $\overline C_1(V)$  &  $\overline L_1(V)$ & $\overline M_1(A)$  &  $\overline E_1(V)$ \\
\hline
 $Q^{-3}$ & $  0.00171$ & $  0.00000$ & $  0.03695$ & $  0.00000$ \\
 $Q^{-2}$ & $  0.00171$ & $  0.00171$ & $  0.03695$ & $  0.00150$ \\
 $Q^{-1}$ & $  0.00327$ & $  0.00328$ & $  0.03696$ & $  0.00267$ \\
 $Q^{0}$  & $  0.00343$ & $  0.00344$ & $  0.03663$ & $  0.00263$ \\
\hline
 & \multicolumn{4}{c}{${}^3P_2$ RMEs}\\
Order & $\overline{C_2}({A})$ & $\overline{L_2}({A})$ & $\overline{M_2}({V})$ & $\overline{E_2}({A})$ \\
\hline
 $Q^{-3}$ & $  0.00000$ & $  0.03387$ & $  0.00000$ & $  0.04226$ \\
 $Q^{-2}$ & $ -0.00058$ & $  0.03387$ & $ -0.00160$ & $  0.04226$ \\
 $Q^{-1}$ & $ -0.00061$ & $  0.03380$ & $ -0.00168$ & $  0.04218$ \\
 $Q^{0}$  & $ -0.00061$ & $  0.03457$ & $ -0.00168$ & $  0.04313$ \\
        \hline
        \hline
\end{tabular}
  \caption{\label{tab:hep3}
    The same as Table~\protect\ref{tab:hep2} but for the ${}^3S_1$, ${}^1P_1$, ${}^3P_1$, and ${}^3P_2$ RMEs
    for the hep process at zero $p- \het$ CM energy.
    Note that the ${}^1P_1$ and ${}^3P_1$ (${}^3S_1$ and ${}^3P_2$) RMEs are purely real (imaginary)
    and in fm$^{3/2}$ units.}
    \end{center}
\end{table}

The S- and P- waves with $J>0$ RMEs are reported in Table~\ref{tab:hep3}. 
The $^3S_1$ capture is induced by the weak axial charge and current,
and weak vector current operators via the multipoles 
$C_1({A})$, $L_1({A})$, $E_1({A})$, and $M_1({V})$.
Naively, one would expect the RMEs  $L_1({A})$ and $E_1({A})$ to
be dominant, as they take contribution already  at order $Q^{-3}$ by the Gamow-Teller
operator. However, as already discussed, the contribution of this
operator is suppressed at low $q$ values. For that reason, these
RMEs are of the same order of the ${}^3P_0$ RMEs. Moreover,
they take large contributions from the two-body currents at order $Q^{-1}$ and $Q^0$,
which however interfere destructively, leaving the RMEs
to be approximately the same as those calculated at LO. 

The $^1P_1$ and $^3P_1$ captures are induced by the weak vector charge
and current, and weak axial current via the multipoles
$C_1({V})$, $L_1({V})$, $E_1({V})$, 
and $M_1({A})$.
The RME magnitudes of the weak vector transitions in $^3P_1$
capture are much smaller than those in $^1P_1$ capture.  In the long-wavelength
approximation, the one-body $C_1({V})$, $L_1({V})$, and
$E_1({V})$ multipoles are independent of spin, and therefore
cannot connect the dominant part of the $^3P_1$ wave function,
which has total spin $S$=1, to the dominant S-wave component of $^4$He, which
has $S$=0.  This is not the case for the $^1P_1$ channel, which
has total spin $S$=0. 
Morever, due to the suppression of the transition from the $^3P_1$ wave,
the two-body contribution to the $^3P_1$
$C_1({V})$, $L_1({V})$, and $E_1({V})$ RMEs
become relevant.
On the other hand, the $C_1({V})$, $L_1({V})$, and $E_1({V})$ RMEs
from the  $^1P_1$ wave are rather large. 
The situation is reversed for the axial transition, since there
the spin-flip nature of the $M_1({A})$ multipole
makes the associated RME in $^3P_1$ larger than that in
$^1P_1$ (in absolute value).

The $^3P_2$ capture is induced by the weak axial charge and current, 
and weak vector current operators via the multipoles 
$C_2({A})$, $L_2({A})$, $E_2({A})$, and $M_2({V})$. In this case, there is
not any specific suppression of the
RMEs, therefore the $L_2({A})$ and $E_2({A})$ RMEs are large,
and dominated by the contributions of one-body LO currents.
In fact, the latter operator  can now connect the large
S-wave components of both three- and four-nucleon bound states.
The effect of the current operators at higher order is subleading.
Note that these two RMEs are in any case comparable 
to the $L_1({A})$ and $E_1({A})$ RMEs in $^3S_1$ capture. 
The RMEs $C_2({A})$ and $M_2({V})$ are small
because they receive contribution only from subleading current operators.
The RMEs involving partial waves with $L\geq 2$  are very tiny and not worth to be discussed here. In the final calculation
we have taken into account them up to $L=2$.

\begin{figure}[bth]
\centering
\includegraphics[scale=0.40,clip]{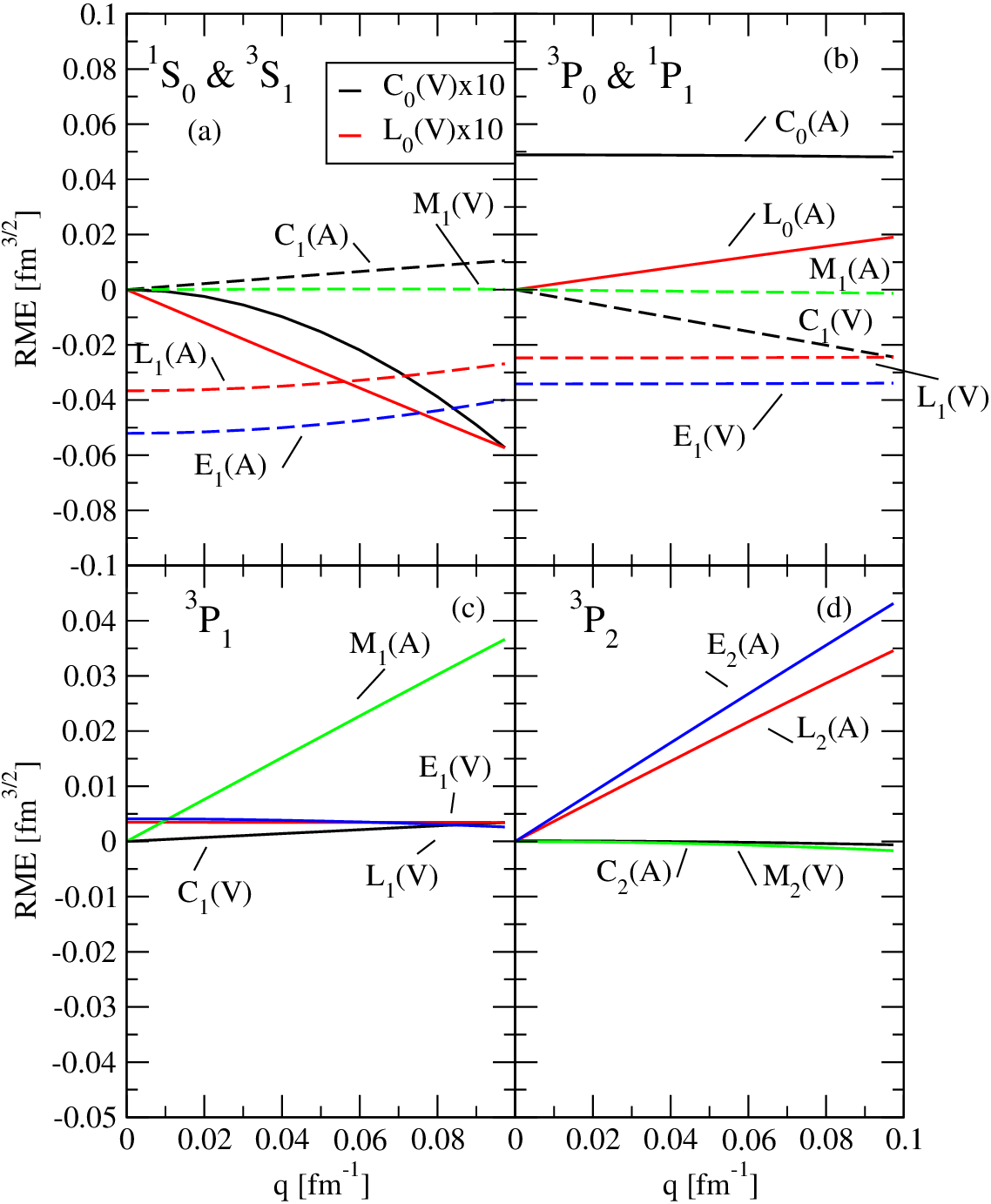}
\caption{(color online)  $q$ dependence of various RMEs for the hep process at zero $p- \het$ CM energy.
  The shown RMEs are all rescaled as in Eq.~(\ref{newRME2}).
    The results have been obtained with the N4LO500/3N interaction and the N3LO accompanying weak current. 
    Note that the ${}^1S_0$, ${}^1P_1$, and ${}^3P_1$ (${}^3P_0$, ${}^3S_1$ and ${}^3P_2$) RMEs are purely real
    (imaginary) and in fm$^{3/2}$ units.}
  \label{fig:rme}
\end{figure}

In Fig.~\ref{fig:rme} we report the behavior of the
most relevant RMEs as function of the momentum transfer $q$. The dependence is of polynomial type, and can be fitted with  a  second order polynomial. For example, the RME $C_0^{000}(V)$ displayed as a black line in panel (a) of Fig.~\ref{fig:rme} is proportional essentially to $q^2$. This behavior can be easily
understood by considering the matrix element of the vector charge operator at LO, namely the Fermi operator,
\begin{equation}
  \langle \heq | \sum_{j=1,4} e^{i\bmq\cdot\bmr^{(CM)}_j} \tau_-(j)| p+\het, {}^{1}S_0\rangle \ ,
\end{equation}
where $\bmr^{(CM)}_j$ above is the distance of particle $j$ to the CM of the four particle system (the dependence
on the CM position $R_{CM}$ has been integrated out to obtain the momentum conservation), and $\tau_-(j)$ is the ladder isospin operator acting on particle $j$. Expanding the exponential function,
the odd powers of $(i\bmq\cdot\bmr^{(CM)}_j)$ vanish due to the parity constraint. The constant term 
also vanishes since in the integral 
\begin{equation}
  \langle \heq | \sum_{j=1,4} \tau_-(j) | p+\het, {}^{1}S_0\rangle \ ,
\end{equation}
we have $\sum_j \tau_-(j)= 2T_-$, where $\bm{T}$ is the total isospin operator. 
The $\heq$ state is in very good approximation a state of total isospin $T=0$, while the $p+\het$ state is
of total isospin $T=1$. Since $T_-$ cannot change the total isospin, this matrix element is only proportional
to the very tiny $T=1$ components of $\heq$ and this contribution  vanishes due to the fact that
bound and scattering states are mutually orthogonal. Therefore, the leading
contribution to $C_0(V)$ is proportional to $q^2$.
Due to the relation given in Eq.~(\ref{eq:c0l0}), then $L^{000}_0(V)~\sim q$ as clearly shown by the red solid line in panel (a) of Fig.~\ref{fig:rme}. Moreover for $q\approx 0.1$ fm${}^{-1}$ the values of $L^{000}_0(V)$ and $C_0^{000}(V)$ are
almost coincident, as predicted by Eq.~(\ref{eq:c0l0}) since $E_3-E_4\approx 0.1$ fm${}^{-1}$. The same
analysis discussion is valid  for the RMEs $C_1(V)$ and $L_1(V)$ of panel (b).

Panel (c) shows clearly the suppression of most of the RMEs for the ${}^3P_1$ wave, only
$M_1(A)$ is sizable. In panel (d), we can notice that the RME $C_2(A)$ is very suppressed.
The leading contribution of this RME comes from the matrix of the NLO operator given in Eq.~(4.10) of Ref.~\cite{Marcucci:2001dmc}, i.e. it is proportional to
\begin{equation}
  \langle \heq | \biggl( \sum_{j=1,4} \bmsi_j\cdot [ \bmp_j,e^{i\bmq\cdot\bmr^{(CM)}_j}]_+ \tau_{-}(j)\biggr) | p+\het, {}^{3}P_2\rangle \ ,
\end{equation}
where $[\cdots , \cdots]_+$ denotes the anticommutator operator, $\bmsi_j$ is the spin operator, and $\bmp_j$ the momentum of nucleon $j$.
Expanding the exponential, the first term gives a vanishing contribution since
the scalar operator $\bmsi_j\cdot\bmp_j$ 
cannot connect the $J=0$ $^4$He and the $^3P_2$ $p+\het$ states,
while the odd-power terms vanish due to parity.
Consequently the first nonvanishing contribution is proportional to $q^2$.

\begin{figure}[bth]
\centering
\includegraphics[scale=0.40,clip]{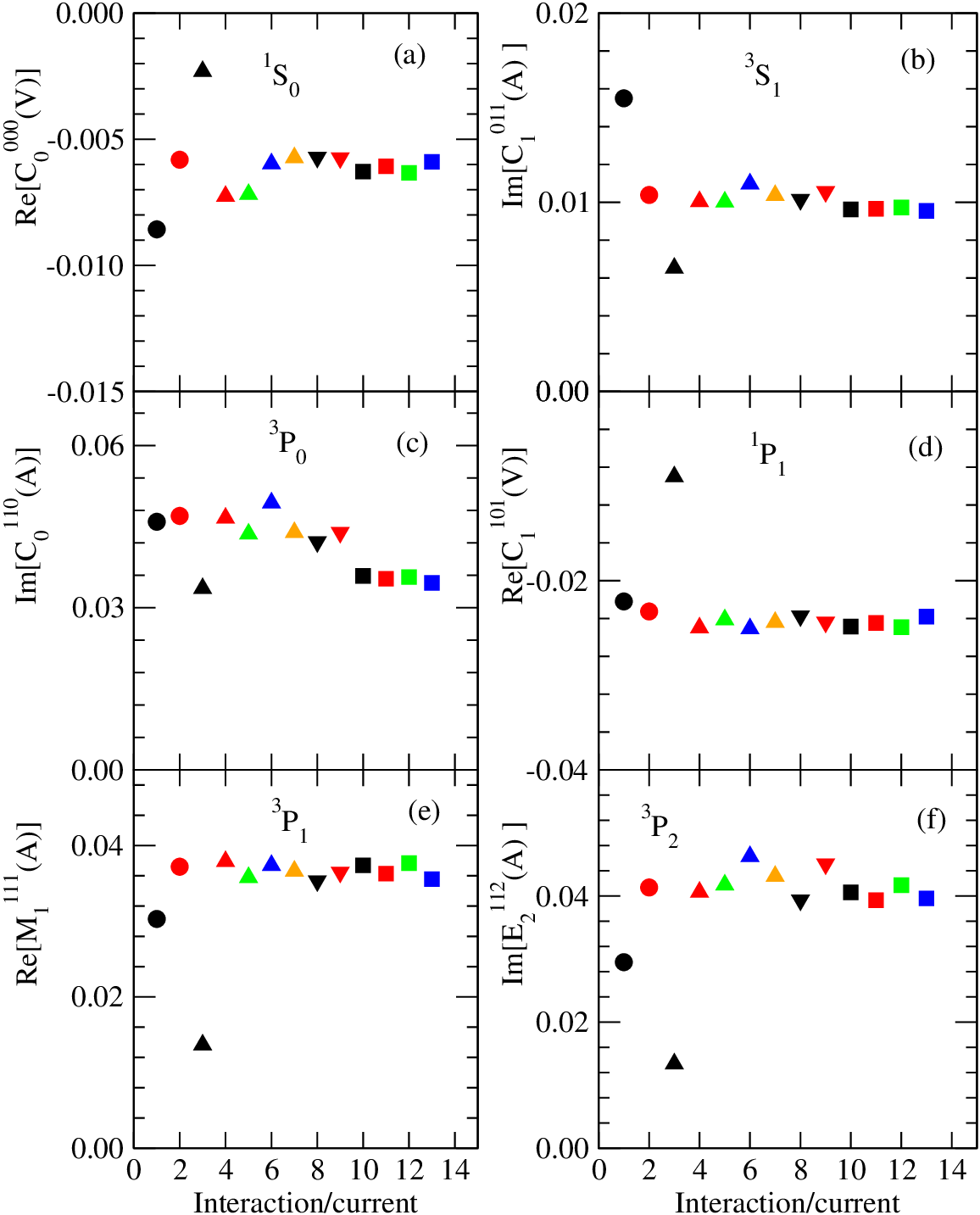}
\caption{(color online) Selected set of RMEs calculated using the various interactions and currents employed in this work.
  The shown RMEs are all rescaled as in Eq.~\eqref{newRME2}.
  The interactions/currents (IC) are numbered from 1 to 13, as explained below.
  IC=1 (black solid circles) results from Ref.~\protect\cite{Marcucci:2001dmc} obtained with the AV18/UIX interaction and the ``full'' model for the weak currents. Note that the old CHH wave functions have been used.
  IC=2 (red solid circles) results obtained using the AV18/UIX interaction in conjunction with the N3LO chiral currents, and the ``modern'' HH wave functions.
  For the rest of the cases IC=3-13 we have used  the N3LO chiral currents and the ``modern'' HH wave functions.
  IC=3-7 (black, red, green, blue, and orange solid up triangles) LO500, NLO500, N2LO500/3N, N3LO500/3N, N4LO500/3N interactions, respectively.
  IC=8,9 (black and red solid down triangles)  N4LO450/3N and N4LO550/3N interactions, respectively.
  IC=10-13 (black, red, green, and blue solid squares)  NVIa, NVIb, NVIIa, and NVIIb interactions, respectively.
  All RMEs are in units of fm${}^{3/2}$.
  }
  \label{fig:rme_comp}
\end{figure}

In Fig.~\ref{fig:rme_comp} we report a selected number of RMEs obtained with the various interactions and currents employed in this work,
comparing them with those obtained in  Ref.~\cite{Marcucci:2001dmc}. 
First of all, we compare the old and new calculations performed with the AV18/UIX interaction. The old results (solid black circles)
were obtained in Ref.~\cite{Marcucci:2001dmc} using the CHH wave functions and phenomenological currents. The new
results (solid red circles) have been obtained with the same interaction, but the wave functions were obtained using the
more accurate HH method and the chiral weak current described in Sec.~\ref{sec:models}. As it can be seen by
inspecting Fig.~\ref{fig:rme_comp}, we observe some sizable difference in the two cases, in particular for the
$C_0^{000}(V)$, $C_1^{011}(A)$, $M_1^{111}(A)$, and $E_2^{112}(A)$ RMEs. In the other cases, the two calculations give RMEs
rather close. While it is difficult to disentangle the origin of the differences, the most probable reason is the lower quality of the
CHH wave functions used in Ref.~\cite{Marcucci:2001dmc}.
In any case, we anticipate that the final S-factor values are not so different for the two calculations.

The RMEs values shown in Fig.~\ref{fig:rme_comp} as black, red, green, blue, and orange solid up triangles correspond to the calculations
performed using the wave functions calculated with the LO500, NLO500, N2LO500/3N, N3LO500/3N, and N4LO500/3N
interactions, respectively. In all cases, we have considered the current at N3LO.
In this way, it is possible to explore the dependence on the chiral order of the
nuclear interaction. As it can be seen, the RMEs calculated with the LO500 interaction
are rather at variance with respect to the other cases. This is related to the fact that
with this interaction, the $\het$ and $\heq$ binding energies are  overpredicted
with respect to the experimental ones (see Table~\ref{tab:potlist}). For the other interactions,
the RMEs are quite close, in particular for the N3LO500/3N and N4LO500/3N cases (blue and orange solid up triangles).

The RMEs shown as black and red solid down triangles are calculated with the N4LO450/3N and
N4LO550/3N interactions. Together with the solid orange up triangle (calculated with the N4LO500/3N interaction),
they show the dependence on the cutoff $\Lambda$ used to regularize the potential and current. In particular for the RMEs receiving contribution from the axial charge and current the three results are somewhat different. 
Finally, the black, red, green, and blue squares correspond to the RMEs calculated with the NV potentials.
Also in this case, the values of the RMEs are rather close to each other. We note, however, a 
shift of values with respect to the RMEs calculated with the EMN interactions, in particular for the $C_0^{110}(A)$ multipole. Similar results are found for the RMEs not shown in Fig.~\ref{fig:rme_comp}.

\subsection{hep at zero energy: S-factor}
\label{subsec:hep0S}
The cumulative contributions to the S-factor adding the various waves in the scattering
wave function is reported in Table~\ref{tab:sfactor1}. The results are shown
for the N4LO500/3N and NVIa/3N interactions and accompanying N3LO chiral current at zero $p+\het$ CM energy.
 As it can be seen by inspecting  Table~\ref{tab:sfactor1},
the wave ${}^3S_1$ brings approximately half of the total contribution. Adding the ${}^1S_0$ wave has a very tiny effect.
The contribution of the ${}^3P_1$ wave is around $0.5 \times 10^{-20}$ keV b in both cases, while
those of the ${}^1P_1$ are  more relevant, being of the order of $\sim 1\times 10^{-20}$ keV b. The main difference in the two calculations is the contribution
of the ${}^3P_0$ wave. In fact, for the NV interactions it results smaller
than that calculated with the EMN interactions. 
As discussed above, the ${}^3P_0$ contribution comes from a destructive interference
between the $C_0^{110}(A)$ and $L_0^{110}(A)$ RMEs, which results  more relevant for the NV interactions. Finally, the contribution of the ${}^3P_2$  wave is rather
sizable, being of the order of $2 \times 10^{-20}$ keV b. The contribution of $D$ and  higher waves is
negligible, as it will be explicitly shown in the next subsection.
Similar results are observed for the other EMN and NV interactions as well.

\begin{table}[t]
\begin{center}
\begin{tabular}{lcc}
\hline 
\hline
Wave & N4LO500/3N & NVIa/3N \\
\hline
${}^3S_1$ & $4.53$ & $4.45$ \\
$+{}^1S_0$ & $4.54$ & $4.46$ \\
$+{}^3P_0$ & $5.56$ & $4.79$ \\
$+{}^1P_1$ & $6.78$ & $5.96$ \\
$+{}^3P_1$ & $7.26$ & $6.47$ \\
$+{}^3P_2$ & $9.42$ & $8.48$ \\
\hline
\hline
\end{tabular}
  \caption{\label{tab:sfactor1}
    Cumulative hep S-factor, in units of $10^{-20}$ keV b, calculated with N4LO500/3N and NVIa/3N interactions
    and accompanying N3LO chiral currents at zero $p+\het$ CM energy, adding in the scattering
    wave function the specified waves. }
    \end{center}
\end{table}

The calculated S-factor for all the considered interactions is presented in Table~\ref{tab:sfactor2},
for the different choices of the weak current chiral orders. The results in the last column, i.e.\ using
the full N3LO weak current, cluster around the value $S(0)=(8-10)\times 10^{-20}$ keV b, quite in agreement
with the old value of Ref.~\cite{Marcucci:2001dmc}, reported in the first row of the table. 
It is worth noticing that the NV family of interactions predicts in general a value of $S(0)$ smaller than the EMN family. A similar systematic difference on the electroweak observables among the two families of interactions has been observed in other recent works that studied the muon capture~\cite{Gnech:2023mvb} and in a recent work on the $pp$ reaction~\cite{Barlucchi:2026}. This last work traced back the difference to cutoff effects generated in the deuteron wave function. Similar effects might be present also in the present bound-state wave functions.

\begin{table}[t]
\begin{center}
\begin{tabular}{lcccc}
\hline 
\hline
Interaction & LO  &  NLO & N2LO & N3LO   \\
\hline
AV18/UIX \protect\cite{Marcucci:2001dmc} & \multicolumn{4}{c}
{$9.64$} \\
\hline
AV18/UIX   & $9.18$ & $10.79$ & $15.79$  & $\n9.47$ \\
LO500      & $4.85$ & $\n5.82$& $\n8.13$ & $\n8.69$ \\
NLO500     & $5.26$ & $\n6.16$& $\n8.40$ & $\n8.53$  \\
N2LO500/3N & $6.75$ & $\n7.72$& $10.19$ & $\n8.18$  \\
N3LO500/3N & $8.43$ & $\n9.86$& $13.13$  & $10.41$  \\
N4LO500/3N & $7.09$ & $\n8.46$& $11.53$  & $\n9.42$  \\
N4LO450/3N & $7.93$ & $\n9.38$& $12.43$  & $\n8.46$ \\
N4LO550/3N & $7.42$ & $\n8.81$& $11.94$  & $\n9.59$  \\
NVIa/3N    & $9.15$ & $10.65$ & $\n6.46$ & $\n8.48$  \\
NVIb/3N    & $9.22$ & $10.77$ & $\n4.76$ & $\n8.18$  \\
NVIIa/3N   & $9.45$ & $10.98$ & $\n6.66$ & $\n8.53$  \\
NVIIb/3N   & $8.62$ & $10.14$ & $\n4.51$ & $\n8.06$  \\
\hline
\hline
\end{tabular}
  \caption{\label{tab:sfactor2}
    The hep S-factor $S(0)$, in units of $10^{-20}$ keV b, calculated with the HH wave functions corresponding to the
    different interactions at zero $p+\het$ CM energy and including all $S+P$ incident channels. 
    The columns labeled LO, NLO, N2LO, and N3LO list the cumulative S-factors
    obtained by retaining components of the specified chiral order in the weak current.
    The value reported in the first row has been obtained in Ref.~\protect\cite{Marcucci:2001dmc}
    with the AV18/UIX interaction and the full model for the phenomenological current.}
    \end{center}
\end{table}

We are now
in the position to try to estimate the theoretical uncertainty coming from the truncation of the chiral expansion of the interaction and the current, and the model dependence. The uncertainties obtained for each interaction are discussed below and summarized in Table~\ref{tab:errors}.
\begin{table}[]
    \centering
    \begin{tabular}{lcccc}
    \hline \hline
        Interaction &  $S(0)$ & $\sigma^I$ & $\sigma^C$ & $\sigma_{\chi {\rm EFT}}$ \\
        \hline
        N4LO450 & 8.46 & 0.19 & 0.64  & 0.83 \\
        N4LO500 & 9.42 & 0.19 & 0.34  & 0.53 \\
        N4LO550 & 9.59 & 0.19 & 0.38  & 0.57 \\
        NVIa    & 8.48 & 0.39 & 0.32  & 0.71 \\
        NVIb    & 8.18 & 0.39 & 0.55  & 0.94 \\
        NVIIa   & 8.53 & 0.39 & 0.30  & 0.69 \\
        NVIIb   & 8.06 & 0.39 & 0.57  & 0.96\\
        \hline \hline
    \end{tabular}
    \caption{The hep S-factor $S(0)$ (in units of $10^{-20}$ keV b)central value for each interaction and relative uncertainty associate to the truncation of the chiral interaction ($\sigma^I$), current ($\sigma^C$) and the sum of the two ($\sigma_{\chi{\rm EFT}}$). See the main text for more details. The values are given in units of $10^{-20}$ keV b.}    \label{tab:errors}
\end{table}

\begin{itemize}
\item {\it Uncertainty related to the truncation of the chiral expansion of the interaction}. For this  we can consider only the $S(0)$ values calculated with the EMN interactions with $\Lambda=500$ MeV
and the corresponding N3LO weak currents for which we have all orders. 
  Using the method of Ref.~\cite{Epelbaum:2014efa} and the values reported in lines 4 to 8 of Table~\ref{tab:sfactor2},
  we compute first the quantities $\Delta\Bigl(S(0)\Bigr)$
  (in units of $10^{-20}$ keV b) via Eqs.~(\ref{eq:dev1})--(\ref{eq:dev4}):
\begin{equation}
  \Delta\Bigl(S(0)\Bigr)=\{0.16,0.35,2.41,1.17\}\,,
\end{equation}
from which we obtain an estimate of  the uncertainties at N3LO and N4LO that reads $\sigma^I_{\rm N3LO}=0.39\times 10^{-20}$, and $\sigma^I_{\rm N4LO}= 0.19\times 10^{-20}$ keV b, respectively. 
Notice that in this case the sequence $\Delta\Bigl(S(0)\Bigr)$ has not the expected behavior, due
to the complex combination of suppressions and cancellations previously discussed.
For example, from Fig.~\ref{fig:rme_comp}, we note that most LO500 RMEs are rather at variance
with those calculated with the EMN interactions at higher orders. However, the LO500 $S(0)$ value turns
out to be very similar to those obtained in the other cases because of the relative cancellations among the various RMEs.
Since we do not have the calculation at all the orders for the remaining interactions, we assume the error $\sigma^I_{\rm N4LO}$ for the EMN N4LO450/3N and N4LO550/3N interactions and the error $\sigma^I_{\rm N3LO}$ for all the NV interactions as shown in Table~\ref{tab:errors}. 

\item  {\it Uncertainty related to
 the truncation of the chiral expansion of the current.} 
 In this case we apply the procedure of Ref.~\cite{Epelbaum:2014efa} to each  nuclear interaction fixed at N4LO for the EMN  models and N3LO for the NV cases. The results are reported in Table~\ref{tab:errors}.
 Looking at the values reported in the
last two columns of Table~\ref{tab:sfactor2}, we note that
 increasing the chiral order of the  weak current from N2LO to N3LO, the change of $S(0)$ is about
$2\times 10^{-20}$ keV b or even greater contradicting the naive chiral expansion convergence. However, we expect that the N3LO values  to be rather accurate, because  the LEC $d_R$ reabsorbs part of the physics missing at the previous orders, being fitted to reproduce the GTME of tritium beta decay. So, the contributions
of the chiral current higher orders should be smaller.
\end{itemize}

\begin{figure}[bth]
\centering
\includegraphics[scale=0.35,clip]{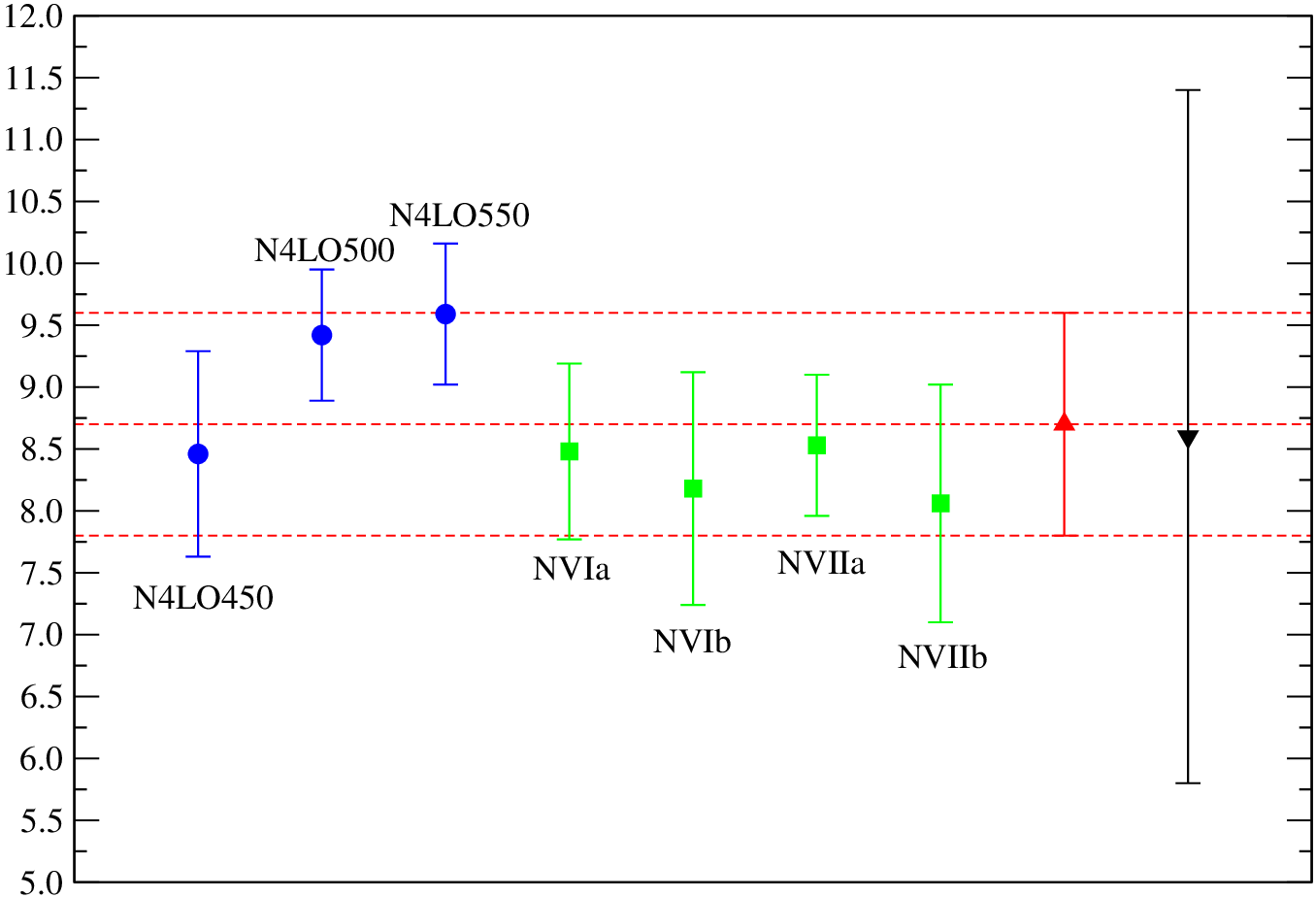}
\caption{(color online) The value of $S(0)$ and relative theoretical uncertainty
(in units of $10^{-20}$ keV b) for the various interaction/current models, calculated
as discussed in the main text. The final suggested value given in Eq.~(\protect\ref{eq:Sfinal})
is also shown (solid red upper triangle), together with the recommended value of Ref.~\cite{SF3} (solid black down triangle). The dashed red lines are guides
for the eyes only.}
  \label{fig:errors}
\end{figure}

Taking into account the values shown in Table~\ref{tab:errors} and using the  model averaging discussed in Sec.~\ref{sec:teoun} our suggested value for the S-factor at zero energy is 
\begin{equation}
    S(0)=(8.7 \pm 0.9)\times 10^{-20}\,\,\text{ keV b}\,,\label{eq:Sfinal}
\end{equation}
that can be compared to that recommended in Ref.~\cite{SF3},
$ S(0)=(8.6 \pm 2.8)\times 10^{-20}$ keV b.
Notice that while the central value is almost identical,  the uncertainty we obtain is three times smaller.
The values for the hep S-factor given in Table~\ref{tab:errors} 
for the various potential/current models and our final suggested value are also shown
in Fig.~\ref{fig:errors} for the sake of clarity.

\begin{figure}[bth]
\centering
\includegraphics[scale=0.35,clip]{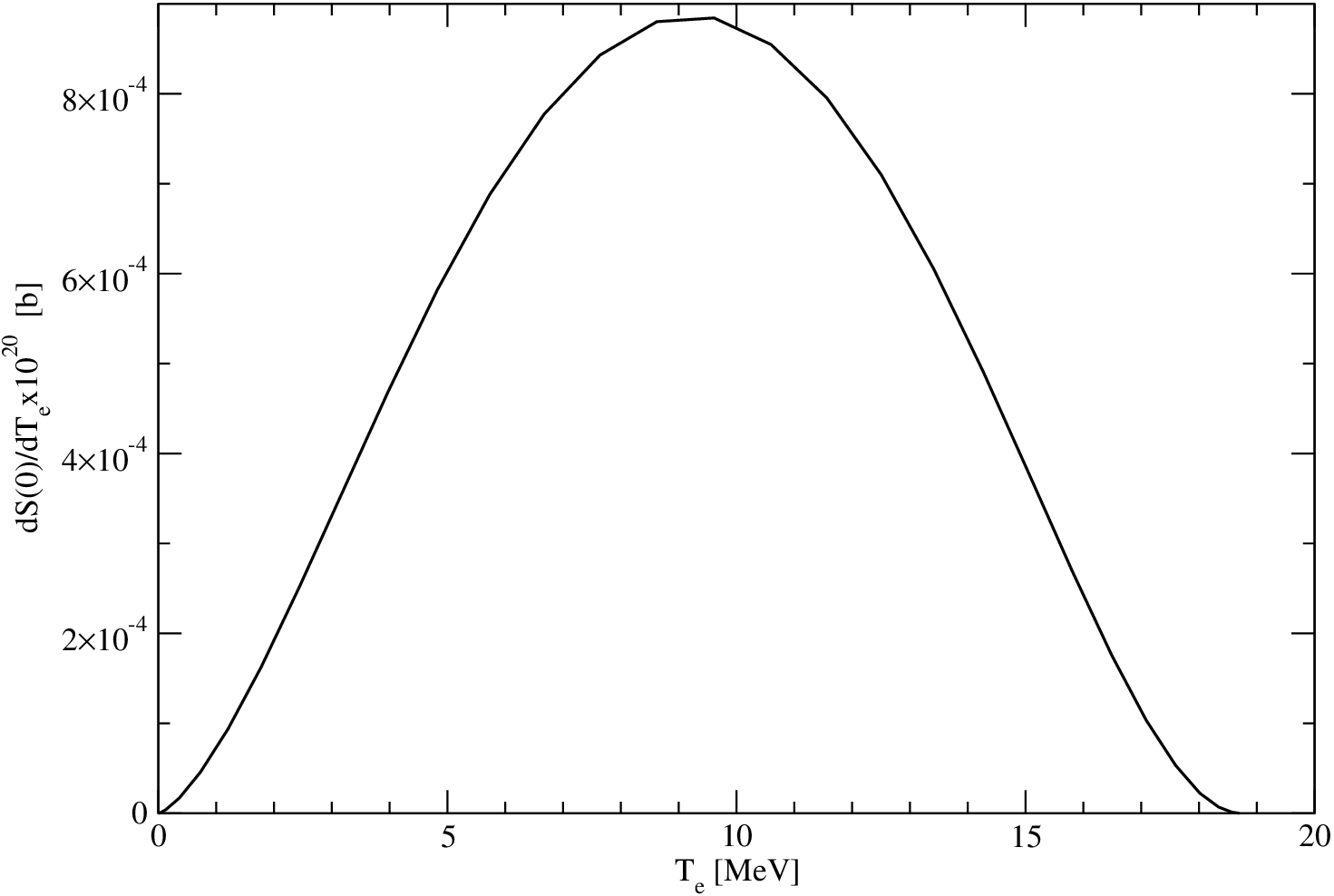}
\caption{(color online) The quantity $dS(0)/dT_e$ (in b) for the hep reaction,
where $T_e$ is the kinetic energy of the outgoing positrons, calculated using
  the EMN N4LO500/3N interaction and accompanying current taken at N3LO. 
  }
  \label{fig:hep_pe}
\end{figure}

Finally, in Fig.~\ref{fig:hep_pe} we show the spectrum of the emitted positrons $dS(0)/dT_e$, $T_e=\sqrt{p_e^2+m_e^2}-m_e$ being the
kinetic energy of the outgoing positrons, calculated for the EMN N4LO500/3N interaction and N3LO accompanying current.
This quantity is proportional to the number of positrons emitted at a given kinetic energy and may be measured in future experiments.

\subsection{hep up to $100$ keV}
\label{subsec:hep100}
\begin{figure}[bth]
\centering
\includegraphics[scale=0.35,clip]{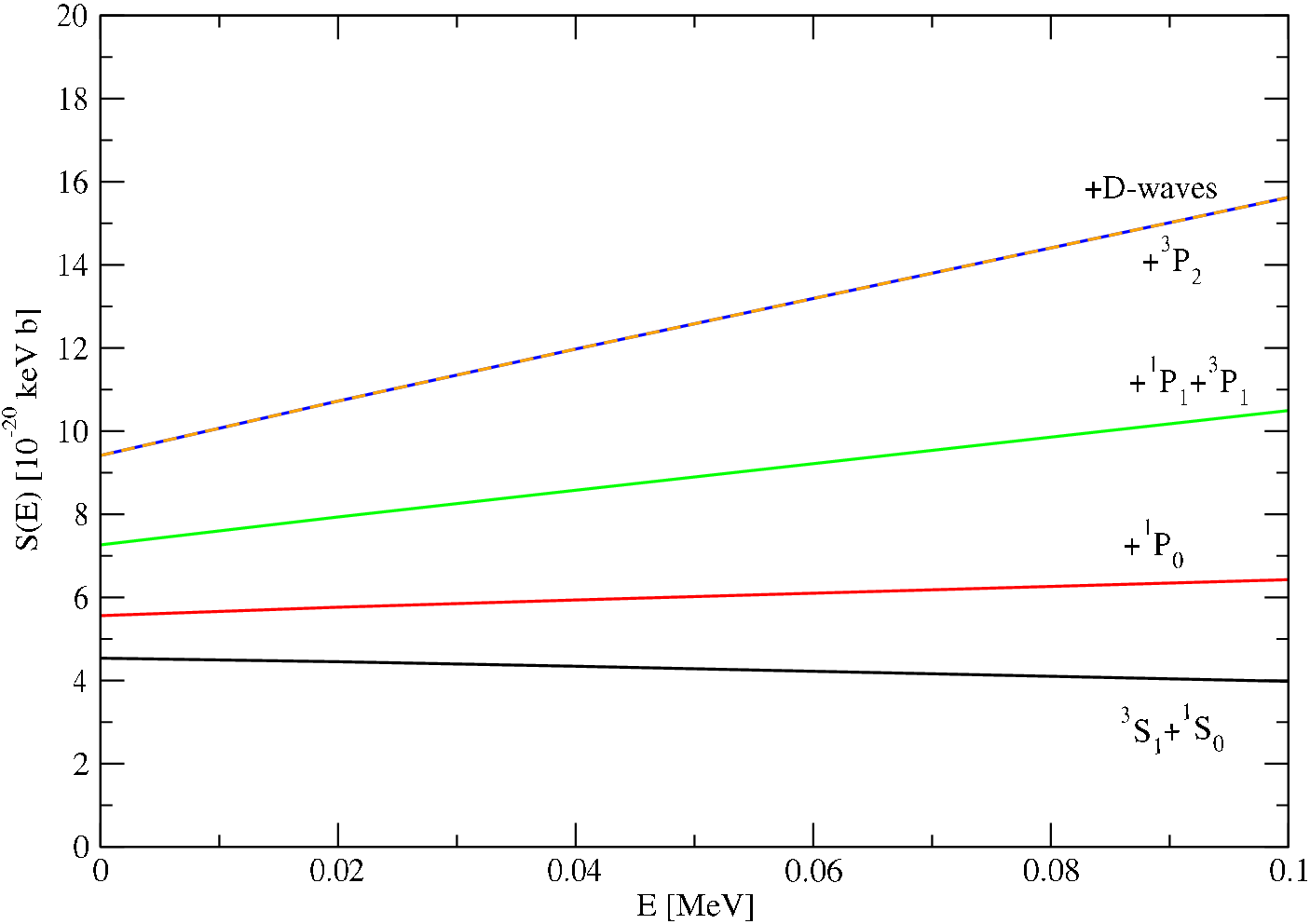}
\caption{(color online) The hep astrophysical factor calculated up to $E_{CM}=0.1$ MeV using
  the EMN N4LO500/3N interaction and accompanying current taken at N3LO. The cumulative contributions
  of the various initial $p+\het$ waves is shown. The effect of the ${}^1S_0$ and of all the D-waves
  (${}^3D_1$, ${}^1D_2$, ${}^3D_2$, and ${}^3D_3$) is very tiny and cannot be distinguished.
  }
  \label{fig:hep100}
\end{figure}

The calculation of the hep astrophysical factor has been extended to CM energies $E$ up to 100 keV. Due to the high computational cost, this study was limited to the N3LO500/3N, N4LO500/3N, NVIa/3N, and NVIb/3N interactions and accompanying currents at N3LO. The cumulative contribution of the various initial $p+\het$ waves  for the N4LO500/3N interaction are shown in
Fig.~\ref{fig:hep100}.
As it can be seen, the contribution of S-waves is slightly decreasing with $E$, while that of the P-waves
increases linearly with energy. The effect of ${}^1S_0$ and of all D-waves 
(${}^3D_1$, ${}^1D_2$, ${}^3D_2$, and ${}^3D_3$) is negligible over the energy range.
For example, the S-factor at $E=100$ keV taking into account S-, P-, and D-waves is $S=15.6231\times 10^{-20}$ keV b,
almost identical to the value $S=15.6228\times 10^{-20}$ keV b obtained including only all S- and P-waves. Moreover, the contribution of D-waves 
at lower energies is more and more suppressed, and therefore it can be safely neglected. In the same way, we expect 
the contribution of higher waves to be even more suppressed, due to the more and more increasing centrifugal barrier. 

In Fig.~\ref{fig:hep100} it is also possible to notice that the total S-factor (solid purple line)  increases linearly with energy. Therefore, it can be parametrized as
\begin{equation}
  S(E)=S(0)+S'(0) E\ , \label{eq:hep100}
\end{equation}
where $E$ is the CM energy in keV.
Performing the calculation for
the four interactions mentioned above, with the corresponding N3LO currents, we  estimate
a value for the first derivative of the S-factor of
\begin{equation}
    S'(0)=(0.06\pm 0.01)\times 10^{-20} \,\,\text{b}\ ,
\end{equation}
where the theoretical uncertainty is estimated by taking the half width of the interval between the upper and lower value of $S'(0)$. 
It should be remarked that to the best of our knowledge, this is the first estimate of
$S'(0)$ ever made.

\subsection{hen at thermal energies}
\label{subsec:hen0}
Let us now consider  the hen process, i.e.\ the reaction $n(\het,\gamma)\heq$, at neutron (laboratory)
thermal energies ($E_n=0.0253$ eV), for which
accurate measurements of the total cross section exist~\cite{Wolfs:1989vw,Wervelman:1991aax}. In this case, the transition is induced
by the EM current, whose isovector part is related to the vector weak current by a rotation
in the isospin space.
Since we are considering this process at very low energies, the cross section is
dominated by the capture in S-waves. For real photons, only transverse multipoles contribute,
therefore the process essentially proceeds via capture from the ${}^3S_1$ wave, and the RME which
comes into play is an $M_1({\rm EM})$. It is very important to notice that in the hep process, the $M_1(V)$ contribution is
very small (see Table~\ref{tab:hep2}), and therefore the study of the hen process at thermal energies is not
a stringent test of the hep calculation. It remains though an important test of the
quality of the wave functions and vector part of the currents.

It is well known that the $M_1({\rm EM})$ RME is rather sensitive to the contributions
of meson-exchange currents in the EM operator, i.e. to the components beyond LO~\cite{Girlanda:2010vm}. For that reason
we need to consider the EM current up to N4LO (namely, at order $Q^1$), including also the
contribution of pion loops and of several contact diagrams. The full EM current operators are derived in Refs.~\cite{Pastore:2009is,Pastore:2011ip,Piarulli:2012bn}. In the N4LO terms, five unconstrained LECs appear.
These LECs have been fixed in order to reproduce the magnetic moments of $A=2,3$ nuclei, and the threshold deuteron electro-disintegration at backward angles~\cite{Gnech:2022vwr}. Therefore, the value of $M_1({\rm EM})$ is directly constrained by experimental observables through the LECs appearing in the EM current at N4LO. However, we are not able to accurately describe the hen experimental values as we will present later.

Furthermore, the $M_1({\rm EM})$ RME for the hen process is also very sensitive to the initial and
final wave functions, since the contribution of the large S-wave components in the $\heq$ bound-state
and $n+\het$ scattering-state wave functions is suppressed~\cite{Girlanda:2010vm}. This is the reason of the exceptionally
small value of the thermal cross section ($\sim 55$ $\mu$b) with respect to the $n+p$ and $n+d$ radiative
capture cross sections ($\sim330$ mb and $\sim50$ mb, respectively). The final cross section value, therefore, is sensitive to the "small components" beyond  S-waves.
of the initial and final wave functions, and consequently rather dependent on the
interaction used to compute them.

\begin{table}[t]
\begin{center}
\begin{tabular}{lcccc}
\hline 
\hline
Interaction & NLO & N2LO & N3LO   & N4LO   \\
\hline
AV18/UIX   & $15.3$ & $2.7$ & $0.7$ & $54.5$ \\
N2LO500/3N & $10.9$ & $1.3$ & $0.4$ & $58.1$ \\
N3LO500/3N & $13.1$ & $2.7$ & $0.4$ & $25.4$ \\
N4LO500/3N & $9.1$  & $4.9$ & $1.4$ & $86.7$ \\ 
N4LO450/3N & $13.5$ & $0.2$& $1.7$ & $74.4$ \\
N4LO550/3N & $9.4$  & $7.1$ & $2.6$ & $63.2$ \\
NVIa/3N    & $14.6$ & $1.7$ & $6.0$ & $37.8$ \\
NVIb/3N    & $14.9$ & $5.2$ & $19.0$ & $46.7$  \\
NVIIa/3N   & $14.9$ & $2.0$ & $6.1$  & $36.7$ \\
NVIIb/3N   & $14.1$ & $6.1$ & $21.3$ & $41.3$ \\
\hline
Expt.~\protect\cite{Wolfs:1989vw}           &&&& $54\pm6$ \\
Expt.~\protect\cite{Wervelman:1991aax}      &&&& $55\pm 3$ \\
\hline
\hline
\end{tabular}
  \caption{\label{tab:hen1}
    The hen cross in section in units of $\mu$b, calculated with 
    different interactions at thermal $n+\het$ CM energy.
    The columns list the cumulative cross sections
    obtained by retaining components of the specified chiral order in the EM current from NLO to N4LO.
    The experimental results are from Refs.~\protect\cite{Wolfs:1989vw,Wervelman:1991aax}. }\end{center}
\end{table}

The results obtained for the hen cross section calculated with
different interactions at thermal $n+\het$ laboratory energy are reported
in Table~\ref{tab:hen1}. In the table we report the cumulative results obtained
using the EM current from NLO  up to N4LO~\cite{Pastore:2009is,Pastore:2011ip,Piarulli:2012bn}, see also Table~\ref{tab:currents}.
As it can be seen, the cross section at NLO (which includes NR one-body terms) is of the order of $10$ $\mu$b, approximately 20\% of the experimental value, due to the suppression already mentioned. For this reason the one-pion exchange at N2LO results of the same order as the NLO contribution. However, it turns out to be of opposite sign, giving rise to large cancellations in the cross section. The relativistic corrections at N3LO becomes relevant as well, because of the suppression of the previous orders, in particular for the EMN model. At this order the differences between the NV and the EMN interactions is determined by the presence of the $\Delta$-isobar contributions in the current associated to NV interaction. Note that these contributions seem to be sensitive to the choice of the cutoffs (models a or b). Only adding the N4LO contributions, we are able to reproduce the right order of magnitude of the cross section. This shows how sensitive is this observable to the 5 LECs appearing at this order in the currents. 

The final results (fifth column of Table~\ref{tab:hen1}) show a 
rather large spread, especially for the EMN interactions. On the other hand, considering only the NV interactions, the spread is reduced, being the results
all slightly below the experimental value. 
We decided for this reaction to not perform the analysis of the truncation error because of the rather poor convergence of the chiral expansion. However, we provide a recommended value obtained using Eq.~\eqref{eq:meanO} together with a systematic uncertainty obtained using Eq.~\eqref{eq:systO}, in which we neglect the truncation errors (i.e $\sigma^2_{\chi{\rm EFT}}=0$). Our final result reads
\begin{eqnarray}
    \sigma_{\text{hen}}(0)=(53\pm 18)\,\,\mu{\text b}\,.
\end{eqnarray}
Even if with such a large theoretical error, the agreement with the available experimental data turns out to be quite good.

\subsection{hen up to a  few MeV energy}
\label{subsec:hen1}
In this subsection, we will consider the hen capture for neutron beam energies $E_n$ up to 
a few MeV.
For $E_n>1$ keV, the most important contribution comes from P-waves capture, in particular from the $J^\pi=1^-$ scattering state,
via electric dipole transitions. Consequently, in this energy range there are two RMEs that play a dominant role, $E_1^{101}({\rm EM})$ and $E_1^{111}({\rm EM})$.
Looking at Table~\ref{tab:hep3}, we see that the RME $E_1^{101}(V)$ entering the hep process is large
and sizable. Hence, the analysis of the hen cross section at these energies represents a good test
of the accuracy of the hep calculation.
The capture from S-waves (via the $M_1$ RME) and D-waves
are found subdominant. Note that for the calculation of the electric dipole RMEs, we can also
use the Siegert theorem via the EM charge operator, so these RMEs are much less sensitive to the
beyond LO two-body contributions. 
In this range of energies there exist also quite accurate experimental results, in particular those reported in
Refs.~\cite{Komar:1993zz,Wervelman:1989bbx}. Other measurements of the hen cross section are reported in
Refs.~\cite{Zurmuhle:1963zz,Alfimenkov:1980aaa,Alfimenkov:1979bbb}.

\begin{figure}[bth]
\centering
\includegraphics[scale=0.35,clip]{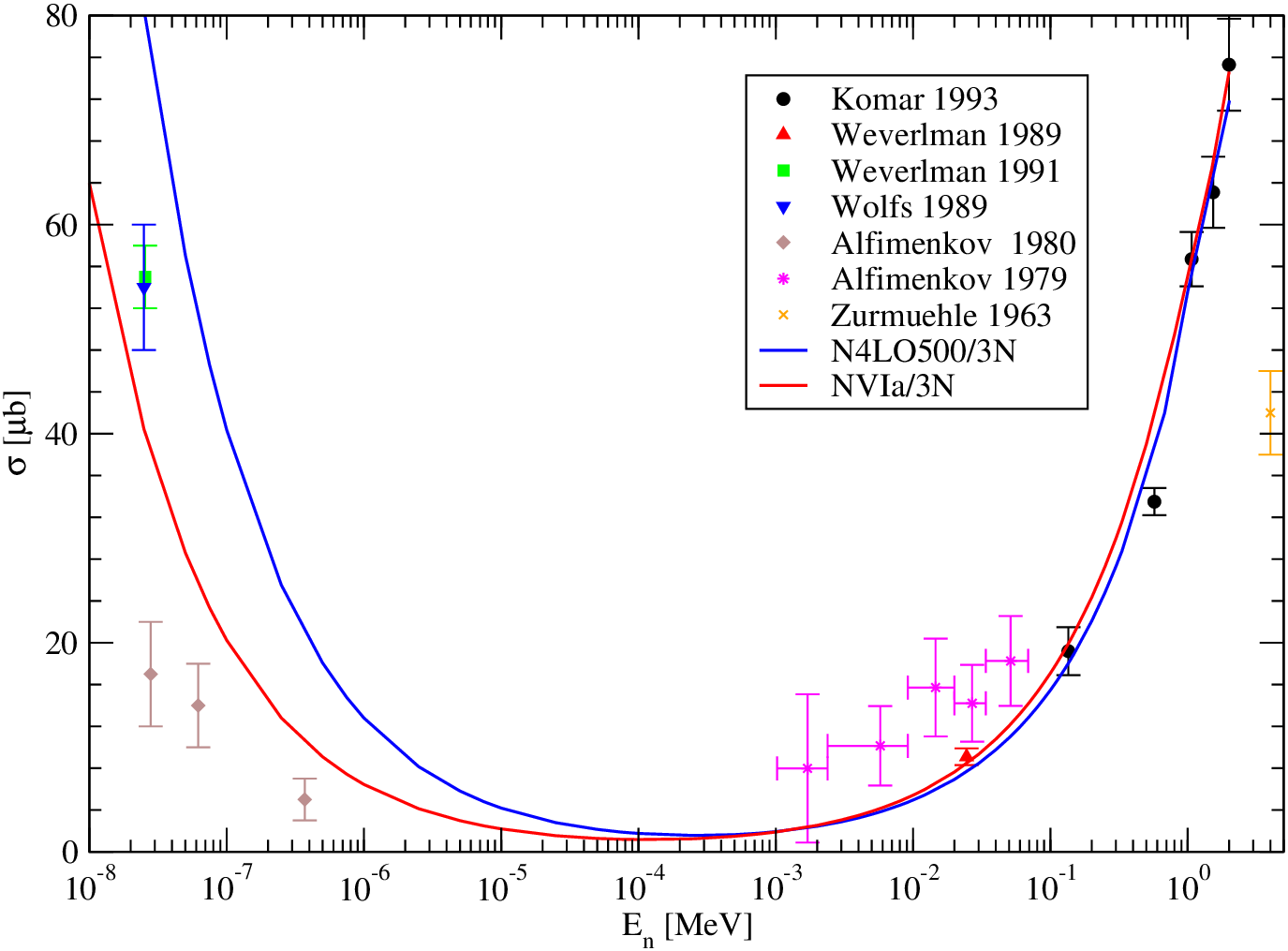}
\caption{(color online) The hen cross section calculated up to a few MeV neutron laboratory energy $E_n$, using
  the  N4LO500/3N and NVIa/3N interactions and accompanying EM currents taken at N4LO, compared
  with the available experimental data.
  }
  \label{fig:sig_log}
\end{figure}

In Fig.~\ref{fig:sig_log}, we show the results obtained with the N4LO500/3N and NVIa/3N interactions and accompanying currents. The $E_1$ RMEs have been obtained 
using the Siegert theorem~\cite{Marcucci:2005zc}.
The large differences at low energies (where the cross section has the well-known $1/v$ law typical of neutron capture)
reflects the difficulty in the calculation of the $M_1^{011}({\rm EM})$ RME discussed in the previous subsection.
Above $1$ keV, P-wave capture starts to become dominant, via the contribution of 
the $E_1^{101}({\rm EM})$ and $E_1^{111}({\rm EM})$ RMEs.
As it can be seen by inspecting the figure, at these energies the different calculations do
not sizably differ and are in good agreement with 
the most recent and accurate measurements reported in Refs.~\cite{Komar:1993zz,Wervelman:1989bbx}. The agreement of the calculation with these data  represents
an optimal proof of the ability of the HH approach together with $\chi$EFT interactions and currents to predict the RME values for these four-nucleon
reactions. There is though some tension between the calculations and the older experimental data of Refs.~\cite{Zurmuhle:1963zz,Alfimenkov:1980aaa,Alfimenkov:1979bbb}. Similar tension can be seen between the new and the old data set. Therefore,  a new set of measurements over this energy region as well as new calculations would help to clarify the situation.

\section{Conclusions}
\label{sec:conc}
In this work, we performed the first refined study of the hep and hen reactions using interactions 
and currents derived within the framework of $\chi$EFT. We
considered two sets of $\chi$EFT interactions, the EMN and NV potentials, together with the 
accompanying currents. An accurate
method based on the HH expansion was used to compute the initial and final wave functions.
We also estimated the theoretical
uncertainty due to the truncation of the chiral expansion of interactions and weak currents,
and computed the uncertainty associated to model dependence.

Using this framework, we performed a detailed analysis of the S-factor of the hep reaction obtaining as final prediction $S(0)=(8.7\pm 0.9)\times 10^{-20}$ keV b, where the error takes into account the theoretical uncertainties arising from chiral order truncation of the interactions and currents, and from model dependence. Notice that our result has a central value very close to the S-factor recommended in Ref.~\cite{SF3}, but with an error three times smaller.
We have also performed a study of  the S-factor energy dependence up to 100 keV, exploring in detail the
importance of D-waves, concluding that they can be safely neglected.  The linear dependence of the S-factor in this energy region allowed us to extract the value of the first derivative at zero energy, found to be
$S'(0)=(0.06\pm0.01)\times10^{-20}$ b. To be remarked that this is the very first estimate for $S'(0)$ ever presented.

As a test of the calculation, we studied in parallel the hen process. At thermal energies,
the calculation is a rather stringent test due to the importance of the beyond LO contributions in
the chiral current and of the small components in the initial and final wave functions. The results obtained for the cross section are found to be of the right order of
magnitude, although quite spread out . 
Our final recommended value  after the analysis of the result is $\sigma_{\text{hen}}(0)=(53\pm18)\,\,\mu$b, in good agreement with the experimental data~\cite{Wolfs:1989vw,Wervelman:1991aax} even if with a very large uncertainty.
Also for this reaction we performed an exploratory study at higher energy up to 
a few MeV.  The calculations well reproduce the increase of the cross section due
to the electric dipole capture, as shown by the most recent data~\cite{Komar:1993zz,Wervelman:1989bbx}. 

It is clear that the current calculations
pass the test provided by the study of the hen reaction giving a strong foundation to the prediction of the hep S-factor. As mentioned before, this work follows all the recommendations mentioned in Ref.~\cite{SF3} for a new analysis of the hep process to obtain a new recommended value.

To further corroborate the validity of the present calculation, we could proceed as follows.
(i) The theory of 3N interaction should be improved by including the N3LO and
N4LO terms. From this point of view, the N4LO 3N interaction includes 13 contact terms, whose LECs
are still to be well determined from 3N data~\cite{Girlanda:2011fh}. (ii) The N4LO weak current (including
the one-pion-loop contribution, 3N terms, and contact interactions) should be
considered. In addition, in this case, some of the new LECs appearing at this order (in the axial charge, in particular)
have yet to be determined. Clearly, in this case the $d_R$ LEC should also be redetermined in order
to reproduce the GTME. (iii) Finally, we could try to perform a Bayesian analysis for the hep and hen observables to determine a statistically meaningful uncertainty. 
All the proposed steps would require a large amount of work, and it is at present unclear whether they will lead to a significant improvement of the present calculation. Nevertheless, work in the proposed 
directions is in progress.

\acknowledgments
The calculations were made possible by grants of computing time
from the Italian National Supercomputing Center CINECA, under the agreement CINECA-INFN,
and from the National Energy Research Scientific Computing Center (NERSC), a Department of Energy User Facility (NERSC award NP-ERCAP 0027079). 
We gratefully acknowledge the support of the INFN-Pisa computing center, as well.
The work of A.G. is supported by the Nuclear Theory for New Physics Topical Collaboration, supported by the U.S.~Department of Energy under contract DE-SC0023663. The support of Jefferson Lab Theory Center, a US Department of Energy facility, under Contract No. DE-AC05-06OR23177 is gratefully acknowledged.

\appendix

\section{Calculation of the wave functions}
\label{app:a}
In this appendix we present the details of the calculation of the bound- and scattering-state wave functions. The HH method adopted here has been described for the $A=4$ systems in several publications~\cite{Kievsky:2008es,Viviani:2020gkm,10.3389/fphy.2020.00069}. Nevertheless, we believe that this appendix may help provide a clearer understanding of the computational aspects of one of the most important ingredients of the present study, namely the initial and final wave functions.

\subsection{The HH expansion basis for bound-state wave functions}
\label{app:bound}

We start with the definition of the Jacobi vectors which,
for a system of four identical particles (disregarding the proton-neutron mass difference), are given by
\begin{eqnarray}
   \jacb_{1p}& = & \sqrt{\frac{3}{2}} 
    \left ({\bm r}_l - \frac{ {\bm r}_i+{\bm r}_j +{\bm r}_k}{3} \right )\ , \nonumber\\
   \jacb_{2p} & = & \sqrt{\frac{4}{3}}
    \left ({\bm r}_k-  \frac{ {\bm r}_i+{\bm r}_j}{2} \right )\ , \label{eq:JcbV}\\
   \jacb_{3p} & =& {\bm r}_j-{\bm r}_i\ , \nonumber
\end{eqnarray}
where $p$ specifies a permutation corresponding to the order $i$, $j$,
$k$ and $l$ of the particles. By definition, the permutation $p=1$ is chosen
to correspond  to the order $1$, $2$, $3$ and $4$. In terms of
the Jacobi vectors, the kinetic energy $T$ is written as
\begin{equation}
  T=-{1\over M}\Bigl( \nabla^2_{\jacb_{1p}}
  +\nabla^2_{\jacb_{2p}}+\nabla^2_{\jacb_{3p}}\biggr)\ ,
\end{equation} 
where $M$ is the nucleon mass.
For a given choice of the Jacobi vectors, the hyperspherical coordinates are
given by the so-called hyperradius $\rho$, defined by
\begin{equation}
   \rho=\sqrt{\jac_{1p}^2+\jac_{2p}^2+\jac_{3p}^2}\ ,\quad ({\rm independent\
    of\ }p)\ ,
    \label{eq:rho}
\end{equation}
and by a set of angular variables which in the Zernike and
Brinkman~\cite{Zernike1935,Ripelle1983} representation are (i) the polar angles
$\hat\jacb_{ip}\equiv (\theta_{ip},\phi_{ip})$  of each Jacobi vector, and (ii) two additional angles, called hyperangles,  $\hypfi_{2p}$ and $\hypfi_{3p}$ defined as
\begin{equation}
    \cos\hypfi_{2p} = \frac{ \jac_{2p} }{\sqrt{\jac_{1p}^2+\jac_{2p}^2}}\ ,
    \quad
    \cos\hypfi_{3p} = \frac{ \jac_{3p} }{\sqrt{\jac_{1p}^2+\jac_{2p}^2+\jac_{3p}^2}}\ ,
     \label{eq:phi}
\end{equation}
where $\jac_{jp}$ is the modulus of the Jacobi vector $\jacb_{jp}$. The set of angular
variables $\hat \jacb_{1p}, \hat \jacb_{2p}, \hat \jacb_{3p}, \hypfi_{2p}$, $\hypfi_{3p}$ is collectively called hyperangles, and is 
denoted  hereafter as $\Omega_p$. The expression of a generic HH
function is
\begin{eqnarray}
 \lefteqn{ {\cal H}^{K,\Lambda, M}_{\ell_1,\ell_2,\ell_3, L_2 ,n_2,
     n_3}(\Omega_p) =\qquad\qquad} &&  \nonumber \\
  && {\cal N}^{\ell_1,\ell_2,\ell_3}_{ n_2, n_3} 
      \left [ \Bigl ( Y_{\ell_1}(\hat \jacb_{1p})
    Y_{\ell_2}(\hat \jacb_{2p}) \Bigr )_{L_2}  Y_{\ell_3}(\hat \jacb_{3p}) \right
    ]_{\Lambda M}  \nonumber \\
  && 
   \times (\sin\hypfi_{2p})^{\ell_1 }    (\cos\hypfi_{2p})^{\ell_2}
      P^{\ell_1+\frac{1}{2}, \ell_2+\frac{1}{2}}_{n_2}(\cos2\hypfi_{2p})
      \nonumber\\
      &&\times
         (\sin\hypfi_{3p})^{\ell_1+\ell_2+2n_2}
      (\cos\hypfi_{3p})^{\ell_3} \nonumber \\
     &&\times  P^{\ell_1+\ell_2+2n_2+2, \ell_3+\frac{1}{2}}_{n_3}(\cos2\hypfi_{3p})\ ,
      \label{eq:hh4P}
\end{eqnarray}
where $P^{a,b}_n$ are Jacobi polynomials and the coefficients ${\cal
N}^{\ell_1,\ell_2,\ell_3}_{ n_2, n_3}$ normalization factors. The quantity 
$K=\ell_1+\ell_2+\ell_3+2(n_2+n_3) $ is the grand angular quantum
number.  The HH functions are the eigenfunctions of the hyperangular part of
the kinetic energy operator. Furthermore, 
$\rho^K   {\cal  H}^{K,\Lambda,M}_{\ell_1,\ell_2,\ell_3, L_2 ,n_2,
n_3}(\Omega_p)$ are homogeneous polynomials of the particle coordinates of
degree $K$.

A set of antisymmetric hyperangular--spin--isospin states of 
grand angular quantum number $K$, total orbital angular momentum $\Lambda$,
total spin $\Sigma$, and total isospin $T$  (for given values of
total angular momentum $J$ and parity $\pi$) can be constructed as follows:
\begin{equation}
  \Psi_{\mu}^{K\Lambda\Sigma T} = \sum_{p=1}^{12}
  \Phi_\mu^{K\Lambda\Sigma T}(i,j,k,l)\ ,
  \label{eq:PSI}
\end{equation}
where the sum is over the $12$ even permutations $p$, and
\begin{eqnarray}
 \lefteqn{  \Phi^{K\Lambda\Sigma T}_{\mu}(i,j,k,l)
   =\qquad\qquad} &&  \nonumber \\
  && \biggl \{
   {\cal H}^{K,\Lambda}_{\ell_1,\ell_2,\ell_3, L_2 ,n_2, n_3}(\Omega_p)
      \biggl [\Bigl[\bigl( s_i s_j \bigr)_{S_a}
      s_k\Bigr]_{S_b} s_l  \biggr]_{\Sigma} \biggr \}_{JJ_z}  \nonumber \\
    && \times  \biggl [\Bigl[\bigl( t_i t_j \bigr)_{T_a}
      t_k\Bigr]_{T_b} t_l  \biggr]_{TT_z}\ .
     \label{eq:PHI}
\end{eqnarray}
Here, ${\cal H}^{K,\Lambda}_{\ell_1,\ell_2,\ell_3, L_2 ,n_2, n_3}(\Omega_p)$ is the
HH state defined in Eq.~(\ref{eq:hh4P}), and $s_i$ ($t_i$) denotes the spin 
(isospin) function of particle $i$. The total orbital angular  momentum $\Lambda$ of
the HH function is coupled to the total spin $\Sigma$ to give the total angular
momentum $JJ_z$, whereas $\pi=(-1)^{\ell_1+\ell_2+\ell_3} $. The
quantum number $T$ specifies the total isospin of the state. The
integer index $\mu$ labels the possible choices of hyperangular, spin and
isospin quantum numbers, namely
\begin{equation}
   \mu \equiv \{ \ell_1,\ell_2,\ell_3, L_2 ,n_2, n_3, S_a,S_b, T_a,T_b
   \}\ ,\label{eq:mu}
\end{equation}
compatibles with the given values of $K$, $\Lambda$, $\Sigma$, 
$T$, $J$ and $\pi$. Each state  $\Psi^{K\Lambda\Sigma T}_\mu$ entering 
the expansion of the four-nucleon wave function must 
be antisymmetric under the exchange of any pair of particles. To this aim 
it is sufficient to consider states such that
\begin{equation}
    \Phi^{K\Lambda\Sigma T}_\mu(i,j,k,l)= 
    -\Phi^{K\Lambda\Sigma T}_\mu(j,i,k,l)\ ,
     \label{eq:exij}
\end{equation}
which is fulfilled when the condition
\begin{equation} 
    \ell_3+S_a+T_a = {\rm odd}\ , \label{eq:lsa}
\end{equation}
is satisfied. Note that many of the antisymmetric states $\Psi^{K\Lambda\Sigma T}_\mu$
are linearly dependent between themselves.

The four-nucleon wave function can be finally written as
\begin{equation}\label{eq:PSI3}
  \Psi_C= \sum_{K\Lambda\Sigma T}\sum_{\mu} 
    u_{K\Lambda\Sigma T\mu}(\rho)
    \Psi_{\mu}^{K\Lambda\Sigma T}\, ,
\end{equation}
where the sum is restricted only to the linearly independent states.  
This expansion can be used to compute either a bound-state wave function or
the ``core'' part of the scattering wave function (see next subsection). We have found convenient to expand the hyperradial functions
$u_{K\Lambda\Sigma T\mu}(\rho)$ in a 
complete set of functions, namely
\begin{equation}
     u_{K\Lambda\Sigma T\mu}(\rho)=\sum_{m=0}^{M-1} 
      c_{K\Lambda\Sigma T\mu m} \; g_m(\rho)
      \ ,     \label{eq:fllag}
\end{equation}
and we have chosen 
\begin{equation}
   g_m(\rho)= 
     \sqrt{b^{9}\frac{m!}{(m+8)!}}\,\,\,  
     L^{(8)}_m(b\rho)\,\,{e}^{-\frac{b}{2}\rho} \ ,
      \label{eq:fllag2}
\end{equation}
where $L^{(8)}_m(b\rho)$ are Laguerre polynomials~\cite{Abramowitz1970} and 
$b$ is a non-linear parameter to be variationally optimized.
One of the problem we have to face is that the number of linearly independent states is still very high, and increases
noticeably with $K$. In order to
reduce the number of states to be included in the expansion, we adopt the same strategy as described in Refs.~\cite{Viviani:2004vf,Viviani:2020gkm}.
Namely, we divide the basis in classes, depending on the value of the quantity $\mathcal{L} =\ell_1 + \ell_2 + \ell_3$
and the values of $n_2$, $n_3$ (see Sec.~3 of Ref.~\cite{Viviani:2004vf} for the detailed definition of the classes).
In this way, we can optimize the number of HH functions needed for expanding the ground state wave functions and
obtaining good and converged values for the $\heq$ ground state binding energy~\cite{Viviani:2004vf}.
See also Ref.~\cite{Duerinck:2025ldb} for an application of this method
to the study of first excited state of $\heq$.

In summary, bound states are calculated using the expansion given in Eq.~(\ref{eq:PSI3}), and expanding the
hyperradial functions as in Eq.~(\ref{eq:fllag}). 
For scattering states, the function given in Eq.~(\ref{eq:PSI3}) is used to describe the core part
of the wave function, namely the region where all the four nucleons are close to each other
(the full wave function will be detailed in the next subsection).

\subsection{The scattering wave function}
\label{sec:scatt}

In the following, a specific clusterization $A+B$ will be
denoted by the index $\gamma$. More specifically, $\gamma=1$, $2$, and $3$ will stand for the
clusterization $p+\tri$, $n+\het$, and $p+\het$, respectively.  Let us consider a scattering state with total
angular momentum quantum number $JJ_z$, and parity $\pi$.
The wave function $\Psi^{\gamma L S}$, describing incoming clusters $\gamma$
with relative orbital angular momentum $L$ and channel spin $S$, coupled to $JJ_z$, can be written as
(the parity $\pi$ being simply $(-)^L$)
\begin{equation}
    \Psi^{\gamma, LSJJ_z}=\Psi_C^{\gamma, LSJJ_z}+\Psi_A^{\gamma, LSJJ_z} \ ,
    \label{eq:psica}
\end{equation}
where the core part $\Psi_C^{\gamma, LSJJ_z}$ describes the four particles when they
are close to each other; it can be conveniently expanded as in Eq.~(\ref{eq:PSI3}).
The other term, $\Psi_A^{\gamma, LSJJ_z}$, describes the relative motion of the two clusters in
the asymptotic regions, where the mutual interaction is
negligible (except for the long-range Coulomb interaction), and it can be decomposed
as a linear combination of the following functions
\begin{eqnarray}
  \Omega_{\gamma, LSJJ_z}^F &=& {1\over\sqrt{4}} {\cal A}\bigg\{
  \Bigl [ Y_{L}(\hat{\bm y}_\gamma)   [ \Phi_\gamma(ijk)  s_l]_{S} 
    \Bigr ]_{JJ_z} \nonumber\\
        & \times&\!\!\! {\frac{F_L(\eta_\gamma,q_\gamma y_\gamma)}{q_\gamma y_\gamma}}\bigg\}\ , \label{eq:psiof}  \\
  \Omega_{\gamma, LSJJ_z}^G &=& {1\over\sqrt{4}} {\cal A}\bigg\{
  \Bigl [ Y_{L}(\hat{\bm y}_\gamma)   [ \Phi_\gamma(ijk)  s_l]_{S}
    \Bigr ]_{JJ_z}   \nonumber\\
  &\times &\!\!\!\frac{G_{L}(\eta_\gamma,q_\gamma y_\gamma)}{q_\gamma y_\gamma} (1-e^{-\beta y_\gamma})^{2L+1}\bigg\}   \ ,  \label{eq:psiog}
\end{eqnarray}
where ${\bm y}_\gamma$ is the distance between the CM of clusters $A$
and $B$, $q_\gamma$ is the magnitude of the relative momentum between the
two clusters, and $\Phi_\gamma(ijk)$ are the bound state wave functions (clearly, $\Phi_1\equiv\Phi_\tri$ and
$\Phi_{2,3}\equiv\Phi_\het$). In the present work, the trinucleon bound state wave functions $\Phi_\gamma(ijk)$ (for both $\het$ and $\tri$)
are described using the $A=3$ HH method~\cite{Kievsky:2008es}.
Since these trinucleon states have positive parities (we disregard in the calculation
any parity-violating interaction), also for these state $\pi=(-)^L$.
Moreover, the channel spin $S$ is obtained coupling the angular momentum of the two clusters. In our
case, we have $S=0,1$. The symbol ${\cal A}$ means that the expression
between the curly braces has to be properly antisymmetrized, summing
over the permutations of the particles $(ijk),l$ with $l=1,\ldots,4$
($\Phi_\gamma(ijk)$ are already antisymmetric under the exchange of $ijk$).

The CM kinetic energy $E_\gamma$ in the channel $\gamma$ is defined by the relations
\begin{equation}
  E_T=-B(\tri)+E_1 = -B(\het)+E_2=-B(\het)+E_3\, \label{eq:energy}
\end{equation}
where $E_T$ is the CM energy of the state and $B(\tri)$ and $B(\het)$ the
binding energies of $\tri$ and $\het$, respectively. Depending on the value
of $E_T$, $E_\gamma$ can be either positive or negative. In the present paper, we
are interested in the range of energies $-B(\het)\leq E_T\leq -2B(d)$, where
$B(d)$ is the deuteron binding energy. Namely, we are below the
opening of the $d+d$ channel. Therefore, always $E_\gamma>0$ and we can define the wave number $q_\gamma$ as
\begin{equation}
  E_\gamma={q_\gamma^2\over 2\mu_\gamma}\ , \qquad
  {1\over \mu_\gamma} = {1\over M_A}+{1\over M_B} 
   \ ,\label{eq:tcm}
\end{equation}
and $M_X$ is the mass of the cluster $X$.
Clearly, in the case of a single nucleon $M_X=M$.

In Eqs.~(\ref{eq:psiof}) and~(\ref{eq:psiog}), the functions $F_L$ and
$G_{L}$ describe the asymptotic radial motion of the
clusters $A$ and $B$. If the two clusters are composed of $Z_A$ and $Z_B$
protons, respectively, the parameter $\eta_\gamma$ is defined as
$\eta_\gamma=\mu_\gamma Z_A Z_B e^2/q_\gamma$, where $e^2=1.43997$
MeV fm. The functions $F_L(\eta,qy)$ and $G_{L}(\eta,qy)$ are 
the regular and irregular Coulomb function, respectively. The term
$(1-e^{-\beta y_\gamma})^{2L+1}$ is used to ``regularize'' the 
irregular Coulomb function for $y_\gamma\rightarrow 0$
(see Ref.~\cite{Viviani:2020gkm} for more details), but it does not affect
its long range behavior. The parameter $\beta$ is usually chosen to be $\beta=0.25$ fm${}^{-1}$.
Let us also define
\begin{equation}
  \Omega_{\gamma, LSJJ_z}^\pm = \Omega_{\gamma, LSJJ_z}^G\pm  {\rm i}  \Omega_{\gamma, LSJJ_z}^F\ {\color{blue}{.}}{\color{red}{,}}
  \label{eq:psiom}
\end{equation}
where $\Omega_{\gamma, LSJJ_z}^+$  ($\Omega_{\gamma, LSJJ_z}^-$) describes the
outgoing (ingoing) relative motion of the clusters specified by $\gamma$. In fact, their asymptotic behaviors is given by
\begin{equation}
  G_{L}(\eta,qy)\pm {\rm i} F_L(\eta,qy) \rightarrow
  e^{\pm {\rm i} \bigl (q y-L\pi/2-\eta\ln(2qy)+\sigma_L\bigr ) }\ ,
\end{equation}
where $\sigma_L$ is the Coulomb phase shift.

If one of the clusters is a neutron, as in the $\gamma=2$ case, then $\eta=0$ and
the functions $F_L$ and $G_L$ reduce to
\begin{equation}
 {F_L(\eta,qy)\over qy } \rightarrow j_L(qy)\ , \qquad
 {G_L(\eta,qy)\over qy } \rightarrow - y_L(qy)\ , 
 \label{eq:eta0}
\end{equation}
where $j_L$ and $y_L$ are the regular and irregular spherical Bessel functions
defined, for example, in Ref.~\cite{Abramowitz1970}.
Finally, the general expression of $\Psi_A^{\gamma, LSJJ_z}$ entering
Eq.~(\ref{eq:psica}) is 
\begin{eqnarray}
  \Psi_A^{\gamma, LSJJ_z}&=& \!\!\!
   \Omega_{\gamma, LSJJ_z}^F\!\!\!
  + \!\!\!\sum_{\gamma'=1}^2 \sum_{L'S'} {\cal T}^{\gamma,\gamma'}_{LS;L^\prime S^\prime}\nonumber \\
  &\times& \Omega_{\gamma', L^\prime S^\prime JJ_z}^+  \ , \quad  \gamma=1,2   \label{eq:psia}\\
  \Psi_A^{3, LSJJ_z}&=& \!\!\!
   \Omega_{3, LSJJ_z}^F\!\!\!
  + \!\!\! \sum_{L'S'} {\cal T}^{3,3}_{LS;L^\prime S^\prime}
  \Omega_{3, L^\prime S^\prime JJ_z}^+\  ,   \label{eq:psia3}
\end{eqnarray}
where the parameters ${\cal T}^{\gamma,\gamma'}_{LS;L^\prime  S^\prime}$ are the so-called $T$-matrix elements.  Of course, the sum over
$L^\prime$ and $S^\prime$ is over  all values compatible with the given $J$ and
parity $\pi$.
Note that for $\gamma=1,2$ in Eq.~\eqref{eq:psia} it is necessary to sum over $\gamma'=1,2$, being both channels with the same third component of the total isospin quantum number $T_z=0$.

The scattering wave functions  $ \Psi^{ph,LSJJ_z}$ (corresponding to $\gamma=3$) and $\Psi^{nh,LSJJ_z}$ (corresponding to $\gamma=2$) needed for the computation of the RMEs are
given explicitly as
\begin{eqnarray}
  \Psi^{ph,lSJJ_z}&=& \Psi_C^{\gamma=3,LSJJ_z}+\Psi_A^{\gamma=3,LSJJ_z}\ ,   \label{eq:psi3}\\
  \Psi^{nh,lSJJ_z}&=& \Psi_C^{\gamma=2,LSJJ_z}+\Psi_A^{\gamma=2,LSJJ_z}\ .   \label{eq:psi2}
\end{eqnarray}
To compute the core parts $\Psi_C^{\gamma,LSJJ_z}$ and the $T$-matrix elements ${\cal T}^{\gamma,\gamma'}_{LS;L^\prime S^\prime}$
we use the Kohn variational principle. The HH functions needed for expanding the various  $\Psi_C^{\gamma,LSJJ_z}$
are again subdivided in classes, as discussed in detail in Ref.~\cite{Viviani:2020gkm}, where also the convergence properties
for the $T$-matrix elements are reported. An example of convergence of some RMEs used in this work is shown in
Table~\ref{tab:hep1}.

%\newpage
\bibliographystyle{apsrev4-1}
\bibliography{biblio}{}

@article{Duerinck:2025ldb,
    author = "Duerinck, P. -Y. and Deltuva, A. and Dohet-Eraly, J. and Gattobigio, M. and Kievsky, A. and Lazauskas, R. and Likandrovas, D. and Viviani, M.",
    title = "{Excited state of the {\ensuremath{\alpha}} particle: A benchmark study}",
    doi = "10.1103/jcd4-8jh8",
    journal = "Phys. Rev. C",
    volume = "112",
    number = "4",
    pages = "044001",
    year = "2025"
}

@article{Girlanda:2011fh,
    author = "Girlanda, L. and Kievsky, A. and Viviani, M.",
    title = "{Subleading contributions to the three-nucleon contact interaction}",
    doi = "10.1103/PhysRevC.84.014001",
    journal = "Phys. Rev. C",
    volume = "84",
    number = "1",
    pages = "014001",
    year = "2011",
    note = "[Erratum: Phys.Rev.C 102, 019903 (2020)]"
}

@article{McDonald:1964zz,
    author = "McDonald, D. G. and Haeberli, W. and Morrow, L. W.",
    title = "{Polarization and Cross Section of Protons Scattered by He-3 from 4 to 13 MeV}",
    doi = "10.1103/PhysRev.133.B1178",
    journal = "Phys. Rev.",
    volume = "133",
    pages = "B1178--B1182",
    year = "1964"
}

@article{Alley:1993zza,
    author = "Alley, M. T. and Knutson, L. D.",
    title = "{Spin correlation measurements for p- He-3 elastic scattering between 4.0 and 10.0 MeV}",
    doi = "10.1103/PhysRevC.48.1890",
    journal = "Phys. Rev. C",
    volume = "48",
    pages = "1890--1900",
    year = "1993"
}

@article{Daniels:2010af,
    author = "Daniels, T. V. and Arnold, C. W. and Cesaratto, J. M. and Clegg, T. B. and Couture, A. H. and Karwowski, H. J. and Katabuchi, T.",
    title = "{Spin-Correlation Coefficients and Phase-Shift Analysis for $p+^3$He Elastic Scattering}",
    doi = "10.1103/PhysRevC.82.034002",
    journal = "Phys. Rev. C",
    volume = "82",
    pages = "034002",
    year = "2010"
}

@article{Zurmuhle:1963zz,
    author = "Zurmuhle, R. W. and Stephens, W. E. and Staub, H. H.",
    title = "{Gamma Rays from Neutron Capture in Helium-3 and Deuteron Capture in Deuterium}",
    doi = "10.1103/PhysRev.132.751",
    journal = "Phys. Rev.",
    volume = "132",
    pages = "751--754",
    year = "1963"
}

@article{Alfimenkov:1980aaa,
    author = "Alfimenkov, V. P. and Borzakov, S. B. and Bunatyan, G. G. and Wierzbicki, J. and Pikelner, L. B. and Sharapov, E. I.",
    title = "{Radiative Capture of Thermal Neutrons by 3He}",
    doi = "",
    journal = "Sov. J. Nucl. Phys.",
    volume = "31",
    pages = "10",
    year = "1980"
}

@article{Alfimenkov:1979bbb,
    author = "Alfimenkov, V. P. and Borzakov, S. B. and Wierzbicki, J. and Ovchinnikov O. N. and Pikelner, L. B. and Sharapov, E. I.",
    title = "{Radiative Capture of He3 Neutrons in the Energy Interval 1-70 keV}",
    doi = "",
    journal = "JETP Lett.",
    volume = "29",
    pages = "91",
    year = "1979"
}

@article{Komar:1993zz,
    author = "Komar, R. J. and Mak, H. -B. and Leslie, J. R. and Evans, H. C. and Bonvin, E. and Earle, E. D. and Alexander, T. K.",
    title = "{He-3 (n, gamma) He-4 cross section and the photodisintegration of He-4}",
    doi = "10.1103/PhysRevC.48.2375",
    journal = "Phys. Rev. C",
    volume = "48",
    pages = "2375--2384",
    year = "1993"
}

@article{Wolfs:1989vw,
    author = "Wolfs, F. L. H. and Freedman, S. J. and Nelson, James E. and Dewey, M. S. and Greene, G. L.",
    title = "{Measurement of the He-3 (n, gamma) He-4 cross-section at thermal neutron energies}",
    doi = "10.1103/PhysRevLett.63.2721",
    journal = "Phys. Rev. Lett.",
    volume = "63",
    pages = "2721--2724",
    year = "1989"
}

@article{Wervelman:1991aax,
    author = "Wervelman, R. and Abrahams, K. and Postma, H. and Booten, J. G. L. and Van Hees, A. G. M.",
    title = "{Nucleon capture by 3 He and the production of solar hep-neutrinos: Cross-section measurements and shell-model calculations}",
    doi = "10.1016/0375-9474(91)90287-G",
    journal = "Nucl. Phys. A",
    volume = "526",
    pages = "265--291",
    year = "1991"
}

@article{Wervelman:1989bbx,
    author = "Wervelman, R. and Postma, H. and Abrahams, K. and Stecher-Rasmussen, F. and Davids, G. J. and Bots, G. J. C.",
    title = "{Cross-Section Measurement of the 3He(n,$\gamma$) Reaction at En = 24.5 keV.}",
    doi = "10.13182/NSE89-A23653",
    journal = "Nucl. Sci. Eng.",
    volume = "102",
    pages = "428",
    year = "1989"
}

@article{Epelbaum:2014efa,
    author = "Epelbaum, E. and Krebs, H. and Mei\ss{}ner, U. G.",
    title = "{Improved chiral nucleon-nucleon potential up to next-to-next-to-next-to-leading order}",
    doi = "10.1140/epja/i2015-15053-8",
    journal = "Eur. Phys. J. A",
    volume = "51",
    number = "5",
    pages = "53",
    year = "2015"
}

@article{Girlanda:2010vm,
    author = "Girlanda, L. and Kievsky, A. and Marcucci, L. E. and Pastore, S. and Schiavilla, R. and Viviani, M.",
    title = "{Thermal neutron captures on $d$ and $^3$He}",
    doi = "10.1103/PhysRevLett.105.232502",
    journal = "Phys. Rev. Lett.",
    volume = "105",
    pages = "232502",
    year = "2010"
}

@article{Marcucci:2000bh,
    author = "Marcucci, L. E. and Schiavilla, R. and Viviani, M. and Kievsky, A. and Rosati, S.",
    title = "{Realistic calculation of the HEP astrophysical factor}",
    doi = "10.1103/PhysRevLett.84.5959",
    journal = "Phys. Rev. Lett.",
    volume = "84",
    pages = "5959",
    year = "2000"
}

@article{Marcucci:2001dmc,
    author = "Marcucci, L. E. and Schiavilla, R. and Viviani, M. and Kievsky, A. and Rosati, S. and Beacom, John F.",
    title = "{Weak proton capture on He-3}",
    doi = "10.1103/PhysRevC.63.015801",
    journal = "Phys. Rev. C",
    volume = "63",
    pages = "015801",
    year = "2000"
}

@article{Park:2002yp,
    author = "Park, T. S. and Marcucci, L. E. and Schiavilla, R. and Viviani, M. and Kievsky, A. and Rosati, S. and Kubodera, K. and Min, D. P. and Rho, M.",
    title = "{Parameter free effective field theory calculation for the solar proton fusion and hep processes}",
    doi = "10.1103/PhysRevC.67.055206",
    journal = "Phys. Rev. C",
    volume = "67",
    pages = "055206",
    year = "2003"
}

@article{SNO:2021xpa,
    author = "Albanese, V. and others",
    collaboration = "SNO+",
    title = "{The SNO+ experiment}",
    doi = "10.1088/1748-0221/16/08/P08059",
    journal = "JINST",
    volume = "16",
    number = "08",
    pages = "P08059",
    year = "2021"
}

@article{Schiavilla:1998je,
    author = "Schiavilla, R. and others",
    title = "{Weak capture of protons by protons}",
    doi = "10.1103/PhysRevC.58.1263",
    journal = "Phys. Rev. C",
    volume = "58",
    pages = "1263",
    year = "1998"
}

@article{Viviani:1994pm,
    author = "Viviani, M. and Kievsky, A. and Rosati, S.",
    title = "{Calculations of the alpha particle ground state}",
    reportNumber = "CEBAF-TH-94-17",
    doi = "10.1007/s006010050002",
    journal = "Few Body Syst.",
    volume = "18",
    pages = "25--39",
    year = "1995"
}

@article{Viviani:1998gr,
    author = "Viviani, M. and Rosati, S. and Kievsky, A.",
    title = "{Neutron H-3 and proton He-3 zero energy scattering}",
    doi = "10.1103/PhysRevLett.81.1580",
    journal = "Phys. Rev. Lett.",
    volume = "81",
    pages = "1580--1583",
    year = "1998"
}

@article{Viviani:2021stx,
    author = "Viviani, M. and Filandri, E. and Girlanda, L. and Gustavino, C. and Kievsky, A. and Marcucci, L. E. and Schiavilla, R.",
    title = "{X17 boson and the H3(p,e+e\ensuremath{-})He4 and He3(n,e+e\ensuremath{-})He4 processes: A theoretical analysis}",
    doi = "10.1103/PhysRevC.105.014001",
    journal = "Phys. Rev. C",
    volume = "105",
    number = "1",
    pages = "014001",
    year = "2022"
}

@article{Acharya:2018qzk,
    author = {Acharya, Bijaya and Ekstr\"om, Andreas and Platter, Lucas},
    title = "{Effective-field-theory predictions of the muon-deuteron capture rate}",
    doi = "10.1103/PhysRevC.98.065506",
    journal = "Phys. Rev. C",
    volume = "98",
    number = "6",
    pages = "065506",
    year = "2018"
}

@article{Bonilla:2022otm,
    author = "Bonilla, Jose and Acharya, Bijaya and Platter, Lucas",
    title = "{Muon capture on the deuteron in chiral effective field theory}",
    doi = "10.1103/PhysRevC.107.065502",
    journal = "Phys. Rev. C",
    volume = "107",
    number = "6",
    pages = "065502",
    year = "2023"
}

@article{Ceccarelli:2022cpz,
    author = "Ceccarelli, L. and Gnech, A. and Marcucci, L. E. and Piarulli, M. and Viviani, M.",
    title = "{Muon capture on deuteron using local chiral potentials}",
    doi = "10.3389/fphy.2022.1049919",
    journal = "Front. in Phys.",
    volume = "10",
    pages = "1049919",
    year = "2023"
}

@article{Gnech:2023mvb,
    author = "Gnech, Alex and Marcucci, Laura Elisa and Viviani, Michele",
    title = "{Bayesian analysis of muon capture on the deuteron in chiral effective field theory}",
    doi = "10.1103/PhysRevC.109.035502",
    journal = "Phys. Rev. C",
    volume = "109",
    number = "3",
    pages = "035502",
    year = "2024"
}

@article{Marcucci:2011jm,
    author = "Marcucci, L. E. and Kievsky, A. and Rosati, S. and Schiavilla, R. and Viviani, M.",
    title = "{Chiral effective field theory predictions for muon capture on deuteron and $^3He$}",
    doi = "10.1103/PhysRevLett.108.052502",
    journal = "Phys. Rev. Lett.",
    volume = "108",
    pages = "052502",
    year = "2012",
    note = "[Erratum: Phys.Rev.Lett. 121, 049901 (2018)]"
}

@article{Golak:2016zcw,
    author = "Golak, J. and Skibi\'nski, R. and Wita\l{}a, H. and Topolnicki, K. and Kamada, H. and Nogga, A. and Marcucci, L. E.",
    title = "{Muon capture on $^3$H}",
    doi = "10.1103/PhysRevC.94.034002",
    journal = "Phys. Rev. C",
    volume = "94",
    number = "3",
    pages = "034002",
    year = "2016"
}

@article{Marcucci:2013tda,
    author = "Marcucci, L. E. and Schiavilla, R. and Viviani, M.",
    title = "{Proton-Proton Weak Capture in Chiral Effective Field Theory}",
    doi = "10.1103/PhysRevLett.110.192503",
    journal = "Phys. Rev. Lett.",
    volume = "110",
    number = "19",
    pages = "192503",
    year = "2013",
    note = "[Erratum: Phys. Rev. Lett. {\bf 123}, 019901 (2019)]"
}

@ARTICLE{10.3389/fphy.2020.00069,
    author = "Marcucci, L. E. and Dohet-Eraly, J. and Girlanda, L. and Gnech, A. and Kievsky, A. and Viviani, M.",
    title = "{The Hyperspherical Harmonics method: a tool for testing and improving nuclear interaction models}",
    doi = "10.3389/fphy.2020.00069",
    journal = "Front. in Phys.",
    volume = "8",
    pages = "69",
    year = "2020"
}

@article{Zernike1935,
    author = "Zernike, F. and Brinkman, H.C.",
    title = "",
    eprint = "",
    journal = "Proc. Kon. Ned. Acad. Wensch.",
    volume = "33",
    pages = "3",
    year = "1935"
}

@article{Ripelle1983,
    author = "Fabre de la Ripelle, M.",
    title = "",
    eprint = "",
    journal = " Ann. Phys. (N.Y.)",
    volume = "147",
    pages = "281",
    year = "1983"
}

@article{Entem:2017gor,
      author         = "Entem, D. R. and Machleidt, R. and Nosyk, Y.",
      title          = "{High-quality two-nucleon potentials up to fifth order of
                        the chiral expansion}",
      journal        = "Phys. Rev.",
      volume         = "C96",
      year           = "2017",
      number         = "2",
      pages          = "024004",
      doi            = "10.1103/PhysRevC.96.024004"
}

@article{Pastore:2009is,
    author = "Pastore, S. and Girlanda, L. and Schiavilla, R. and Viviani, M. and Wiringa, R. B.",
    title = "{Electromagnetic Currents and Magnetic Moments in (chi)EFT}",
    doi = "10.1103/PhysRevC.80.034004",
    journal = "Phys. Rev. C",
    volume = "80",
    pages = "034004",
    year = "2009"
}

@article{Piarulli:2012bn,
    author = "Piarulli, M. and Girlanda, L. and Marcucci, L. E. and Pastore, S. and Schiavilla, R. and Viviani, M.",
    title = "{Electromagnetic structure of A = 2 and 3 nuclei in chiral effective field theory}",
    doi = "10.1103/PhysRevC.87.014006",
    journal = "Phys. Rev. C",
    volume = "87",
    number = "1",
    pages = "014006",
    year = "2013"
}

@article{Baroni:2015uza,
      author         = "Baroni, A. and Girlanda, L. and Pastore, S. and
                        Schiavilla, R. and Viviani, M.",
      title          = "{Nuclear Axial Currents in Chiral Effective Field
                        Theory}",
      journal        = "Phys. Rev. C",
      volume         = "93",
      year           = "2016",
      number         = "1",
      pages          = "015501",
      doi            = "10.1103/PhysRevC.93.049902, 10.1103/PhysRevC.93.015501,
                        10.1103/PhysRevC.95.059901",
      note           = "[Erratum: Phys. Rev.C {\bf 95}, 059901 (2017)]"
}

@article{Baroni:2018fdn,
    author = "Baroni, A. and others",
        title = "{Local chiral interactions, the tritium Gamow-Teller matrix element, and the three-nucleon contact term}",
		        doi = "10.1103/PhysRevC.98.044003",
			    journal = "Phys. Rev. C",
			        volume = "98",
				    number = "4",
				        pages = "044003",
					    year = "2018"
					    }

@article{Wiringa:1994wb,
    author = "Wiringa, Robert B. and Stoks, V. G. J. and Schiavilla, R.",
    title = "{An Accurate nucleon-nucleon potential with charge independence breaking}",
    doi = "10.1103/PhysRevC.51.38",
    journal = "Phys. Rev. C",
    volume = "51",
    pages = "38--51",
    year = "1995"
}

@article{Pudliner:1995wk,
    author = "Pudliner, B. S. and Pandharipande, V. R. and Carlson, J. and Wiringa, Robert B.",
    title = "{Quantum Monte Carlo calculations of A \ensuremath{<}= 6 nuclei}",
    doi = "10.1103/PhysRevLett.74.4396",
    journal = "Phys. Rev. Lett.",
    volume = "74",
    pages = "4396--4399",
    year = "1995"
}

@article{Viviani:2004vf,
    author = "Viviani, M. and Kievsky, A. and Rosati, S.",
    title = "{Calculation of the alpha-particle ground state within the hyperspherical harmonic basis}",
    doi = "10.1103/PhysRevC.71.024006",
    journal = "Phys. Rev. C",
    volume = "71",
    pages = "024006",
    year = "2005"
}

@article{Viviani:2020gkm,
  title = {$n+^{3}\mathrm{H},$ $p+^{3}\mathrm{He},$ $p+^{3}\mathrm{H},$ and $n+^{3}\mathrm{He}$ scattering with the hyperspherical harmonic method},
    author = {Viviani, M. and Girlanda, L. and Kievsky, A. and Marcucci, L. E.},
      journal = {Phys. Rev. C},
        volume = {102},
	  issue = {3},
	    pages = {034007},
	      numpages = {32},
	        year = {2020},
		  month = {Sep},
		    publisher = {American Physical Society},
		      doi = {10.1103/PhysRevC.102.034007},
		        url = {https://link.aps.org/doi/10.1103/PhysRevC.102.034007}
			}

@article{Pastore:2011ip,
      author         = "Pastore, S. and Girlanda, L. and Schiavilla, R. and
                        Viviani, M.",
      title          = "{The two-nucleon electromagnetic charge operator in
                        chiral effective field theory ($\chi$EFT) up to one loop}",
      journal        = "Phys. Rev. C",
      volume         = "84",
      year           = "2011",
      pages          = "024001",
      doi            = "10.1103/PhysRevC.84.024001"
}

@article{Bernard:2007sp,
      author         = "Bernard, V. and Epelbaum, E. and Krebs, H. and Mei{\ss}ner,
                        Ulf-G.",
      title          = "{Subleading contributions to the chiral three-nucleon
                        force. I. Long-range terms}",
      journal        = "Phys. Rev.",
      volume         = "C77",
      year           = "2008",
      pages          = "064004",
      doi            = "10.1103/PhysRevC.77.064004"
}

@article{Bernard:2011zr,
      author         = "Bernard, V. and Epelbaum, E. and Krebs, H. and Mei{\ss}ner,
                        Ulf-G.",
      title          = "{Subleading contributions to the chiral three-nucleon
                        force II: Short-range terms and relativistic corrections}",
      journal        = "Phys. Rev.",
      volume         = "C84",
      year           = "2011",
      pages          = "054001",
      doi            = "10.1103/PhysRevC.84.054001"
}

@article{Krebs:2012yv,
      author         = "Krebs, Hermann and Gasparyan, A. and Epelbaum, Evgeny",
      title          = "{Chiral three-nucleon force at N$^4$LO I: Longest-range
                        contributions}",
      journal        = "Phys. Rev. C",
      volume         = "85",
      year           = "2012",
      pages          = "054006",
      doi            = "10.1103/PhysRevC.85.054006"
}

@article{Krebs:2013kha,
      author         = "Krebs, Hermann and Gasparyan, A. and Epelbaum, Evgeny",
      title          = "{Chiral three-nucleon force at N$^4$LO II:
                        Intermediate-range contributions}",
      journal        = "Phys. Rev. C",
      volume         = "87",
      year           = "2013",
      number         = "5",
      pages          = "054007",
      doi            = "10.1103/PhysRevC.87.054007"
      }

@article{Epelbaum:2002vt,
    author = "Epelbaum, E. and Nogga, A. and Gl{\" o}ckle, Walter and Kamada, H. and Mei\ss ner, Ulf-G. and Witala, H.",
        title = "{Three nucleon forces from chiral effective field theory}",
		        doi = "10.1103/PhysRevC.66.064001",
			    journal = "Phys. Rev. C",
			        volume = "66",
				    pages = "064001",
				        year = "2002"
					}

@article{Marcucci:2005zc,
    author = "Marcucci, L.E. and Viviani, M. and Schiavilla, R. and Kievsky, A. and Rosati, S.",
        title = "{Electromagnetic structure of A=2 and 3 nuclei and the nuclear current operator}",
	    eprint = "nucl-th/0502048",
	        archivePrefix = "arXiv",
		    reportNumber = "JLAB-THY-05-300",
		        doi = "10.1103/PhysRevC.72.014001",
			    journal = "Phys. Rev. C",
			        volume = "72",
				    pages = "014001",
				        year = "2005"
					}

@book{Walecka:1995mi,
      author         = "Walecka, J. D.",
      title          = "{Theoretical nuclear and subnuclear physics}",
	          year      = 1995,
		    publisher = "Oxford University Press",
			address   = "New York"
}

@article{Kievsky:2008es,
    author = "Kievsky, A. and Rosati, S. and Viviani, M. and Marcucci, L.E. and Girlanda, L.",
    title = "{A High-precision variational approach to three- and four-nucleon bound and zero-energy scattering states}",
    doi = "10.1088/0954-3899/35/6/063101",
    journal = "J. Phys. G",
    volume = "35",
    pages = "063101",
    year = "2008"
}

@article{Piarulli:2014bda,
    author = "Piarulli, M. and Girlanda, L. and Schiavilla, R. and Navarro P\'erez, R. and Amaro, J. E. and Ruiz Arriola, E.",
    title = "{Minimally nonlocal nucleon-nucleon potentials with chiral two-pion exchange including $\Delta$ resonances}",
    eprint = "1412.6446",
    archivePrefix = "arXiv",
    primaryClass = "nucl-th",
    reportNumber = "JLAB-THY-14-1993",
    doi = "10.1103/PhysRevC.91.024003",
    journal = "Phys. Rev. C",
    volume = "91",
    number = "2",
    pages = "024003",
    year = "2015"
}

@article{Piarulli:2016vel,
    author = "Piarulli, Maria and Girlanda, Luca and Schiavilla, Rocco and Kievsky, Alejandro and Lovato, Alessandro and Marcucci, Laura E. and Pieper, Steven C. and Viviani, Michele and Wiringa, Robert B.",
        title = "{Local chiral potentials with $\Delta$-intermediate states and the structure of light nuclei}",
		        doi = "10.1103/PhysRevC.94.054007",
			    journal = "Phys. Rev. C",
			        volume = "94",
				    number = "5",
				        pages = "054007",
					    year = "2016"
					    }

@book{Abramowitz1970,
      title     = "Handbook of Mathematical Functions",
            author    = "Abramowitz, Milton and Stegun, Irene",
	          year      = 1970,
		        publisher = "Dover Publications, Inc.",
			      address   = "New York"
			          }

@book{Edmond1957,
      title     = "Angular Momentum in Quantum Mechanics",
            author    = "Edmonds, A.R. ",
	          year      = 1957,
		        publisher = "Princeton University Press",
			      address   = "Princeton"
			          }

@article{Melendez:2019izc,
    author = "Melendez, J. A. and Furnstahl, R. J. and Phillips, D. R. and Pratola, M. T. and Wesolowski, S.",
    title = "{Quantifying Correlated Truncation Errors in Effective Field Theory}",
    doi = "10.1103/PhysRevC.100.044001",
    journal = "Phys. Rev. C",
    volume = "100",
    number = "4",
    pages = "044001",
    year = "2019"
}

@article{Barlucchi:2026,
    author = "Barlucchi, Vittorio and Gnech, Alex and Degl'Innocenti, Scilla and Marcucci, Laura Elisa",
    title = "{Bayesian analysis of proton-proton fusion in chiral effective field theory}",
    eprint = "2603.25465",
    archivePrefix = "arXiv",
    primaryClass = "nucl-th",
    month = "3",
    year = "2026"
}

@article{Gnech:2022vwr,
    author = "Gnech, Alex and Schiavilla, Rocco",
    title = "{Magnetic structure of few-nucleon systems at high momentum transfers in a chiral effective field theory approach}",
    doi = "10.1103/PhysRevC.106.044001",
    journal = "Phys. Rev. C",
    volume = "106",
    number = "4",
    pages = "044001",
    year = "2022"
}

@article{SF3,
  title = {Solar fusion III: New data and theory for hydrogen-burning stars},
  author = {Acharya, B. and Aliotta, M. and Balantekin, A. B. and Bemmerer, D. and Bertulani, C. A. and Best, A. and Brune, C. R. and Buompane, R. and Gialanella, L. and Cavanna, F. and Chen, J. W. and Colgan, J. and Czarnecki, A. and Davids, B. and deBoer, R. J. and Delahaye, F. and Depalo, R. and Guglielmetti, A. and Garc\'{\i}a, A. and Robertson, R. G. H. and Gatu Johnson, M. and Gazit, D. and Greife, U. and Guffanti, D. and Hambleton, K. and Haxton, W. C. and Herrera, Y. and Serenelli, A. and Huang, M. and Iliadis, C. and Kravvaris, K. and La Cognata, M. and Langanke, K. and Marcucci, L. E. and Nagayama, T. and Nollett, K. M. and Odell, D. and Orebi Gann, G. D. and Piatti, D. and Pinsonneault, M. and Platter, L. and Rupak, G. and Sferrazza, M. and Sz\"ucs, T. and Tang, X. and Tumino, A. and Villante, F. L. and Walker-Loud, A. and Zhang, X. and Zuber, K.},
  journal = {Rev. Mod. Phys.},
  volume = {97},
  issue = {3},
  pages = {035002},
  numpages = {74},
  year = {2025},
  month = {Sep},
  publisher = {American Physical Society},
  doi = {10.1103/8lm7-gs18},
  url = {https://link.aps.org/doi/10.1103/8lm7-gs18}
}
\end{document}